\documentclass[usegraphicx,useAMS,usenatbib,usefloat]{mn2e}

\usepackage{subfigure} \usepackage[dvips]{graphicx} \usepackage{setspace}
\usepackage{epsfig,natbib}
\usepackage{mathptm,amsmath,amssymb,mathrsfs}

\begin{document} 
\title[Cosmology with the {\em Planck} cluster sample]     
      {Cosmology with the {\em Planck} cluster sample}
\author[J.~Geisb\"{u}sch, M.~P.~Hobson]  
       {J\"{o}rn~Geisb\"{u}sch$^{1,2}$\thanks{email: joern@mrao.cam.ac.uk; joern@cdf.in2p3.fr}, Michael~P.~Hobson$^{1}$\\
       $^{1}$ Astrophysics Group, Cavendish Laboratory, Magingley Road,
      Cambridge CB3 0HE, United Kingdom\\
      $^{2}$ APC, Coll\`{e}ge de France, 11 pl. Marcelin Berthelot, 75231 Paris Cedex 05, France}

\maketitle


\begin{abstract}
It has been long recognised that, besides being a formidable experiment
to observe the primordial CMB anisotropies, Planck will also have the capability to detect galaxy
clusters via their SZ imprint. 
In this paper constraints on cosmological parameters
derivable from the Planck cluster candidate sample are examined for the first time
as a function of
cluster sample selection and purity obtained from realistic simulations of
the microwave sky at the Planck observing frequency bands, observation
process modelling and a cluster extraction pipeline. In
particular, we employ a multi-frequency matched filtering (MFMF) method to
recover clusters from mock simulations
of Planck observations. 
 Obtainable cosmological constraints under realistic assumptions of priors and knowledge
about cluster redshifts are discussed. Just relying on cluster redshift
abundances without making use of recovered cluster fluxes, it is shown that from the
Planck cluster catalogue cosmological constraints comparable to the ones derived
from recent primordial CMB power spectrum measurements can be achieved.
For example, for a concordance $\Lambda$CDM model and a redshift binning of
$\Delta z = 0.1$, the $1\sigma$ uncertainties on the values of $\Omega_m$ and $\sigma_8$ are
$\Delta\Omega_m \approx 0.031$ and $\Delta\sigma_8 \approx 0.014$
 respectively. Furthermore, we find that the constraint of the
matter density depends strongly on the prior which can be imposed on the
Hubble parameter by other observational means.
\end{abstract}


\section{Introduction}
\label{sec:intro}

The cosmological potential of galaxy cluster surveys via the SZ effect
(\cite{1970SZ}; \cite{Sunyaev72CASP} and \cite{Sunyaev80ARA&A}; recent
reviews: \cite{Rephaeli95ARA&A}; \cite{1999Birkinshaw} and \cite{2002Carlstrometal}) has been
advocated by many theoretical papers in recent years (see
e.g. \cite{2000Bartlett}; \cite{2001Bartlett}; \cite{2005CohnKadota}; \cite{2001Haimanetal}; \cite{2004Molnaretal} etc.). Future measurements of the number density and
distribution of clusters will have a profound impact on our understanding of
the nature of the Universe (see e.g. \cite{1999Bahcalletal}; \cite{2003BoehringerSchuecker}; \cite{2005Voit2}). The SZ effect is due to
its redshift independence especially valuable for detecting clusters.
Apart from SZ dedicated surveys (ACT\footnote{\texttt{http://www.hep.upenn.edu/\~{}angelica/act/act.html}}; AMI\footnote{\texttt{http://www.mrao.cam.ac.uk/telescopes/ami/index.html}}; Amiba\footnote{\texttt{http://amiba.asiaa.sinica.edu.tw}}; APEX-SZ\footnote{\texttt{http://bolo.berkeley.edu/apexsz}};  SPT\footnote{\texttt{http://spt.uchicago.edu}}; SZA\footnote{\texttt{http://astro.uchicago.edu/sza}})
which observe fractions of the sky, the Planck surveyor satellite\footnote{\texttt{http://www.rssd.esa.int/Planck}}, which is
scheduled for launch in 2008, will provide detailed full-sky maps at nine
different observing frequencies, ranging from 30 to 857 GHz. It is thus a
particular suitable instrument to detect the thermal SZ effect owing to its
distinct frequency dependence.

Recently several authors have investigated the properties of a Planck SZ
cluster sample by applying different object detection 
 algorithms to simulated Planck channel observations (see
 e.g. \cite{2005Aghanimetal}; \cite{2002Diegoetal}; \cite{2004bHansen};
 \cite{2002Herranzetal}; \cite{2001Kayetal}; \cite{2005Pierpaolietal};
 \cite{2006Schaeferetal}; \cite{2006SchaeferBartelmann}; \cite{2003White}).
For example, \cite{2002Diegoetal} developed a Bayesian non-parametric method
 to detect clusters in Planck data, which combines Planck frequency channels
 in such a way that the signal of contaminating components is reduced with
 respect to the cluster SZ one. Clusters are then extracted from the resulting map by
 employing SExtractor (\cite{1996BertinArnouts}).
\cite{2002Herranzetal} and \cite{2006SchaeferBartelmann} implemented matched
and scale adaptive filtering (\cite{2001Sanzetal}) techniques to recover galaxy
 clusters and their photometry from Planck multi-frequency observations.
While \cite{2002Herranzetal} work in the Fourier domain and apply the
filters to a $12.8\times12.8$ deg$^2$ sky patch, \cite{2006SchaeferBartelmann}
work in spherical harmonic space and apply the scale adaptive and matched filtering
technique to full-sky Planck simulations using the HEALPIX pixelisation scheme
(\cite{2005Gorskietal}) to store the data. Also \cite{2003SchulzWhite} use a simple
matched filtering algorithm to extract clusters from a map resulting from a
 hypothesised combination of Planck frequency channels.
\cite{2005Pierpaolietal} discusses a wavelet based method for component separation designed to recover non-Gaussian, spatially localized and sparse signals.
A comparison of the selection and contamination of wavelet and matched
filtering extraction techniques is given in \cite{2006ValeWhite}.
In \cite{2005Geisbueschetal} (hereafter GKH05), we applied a cluster extraction algorithm that
combines the Harmonic Space Maximum Entropy Method (HSMEM; see also \cite{2002Stolyarovetal}) with a Peak Finding Flux
Integration method to recover galaxy clusters from realistic full-sky Planck
simulations based on the HEALPIX pixelisation scheme.
Furthermore, there have been purely theoretical efforts based on the cluster mass
function to estimate the power of the
Planck cluster catalogue to constrain cosmological model parameters (see
e.g. \cite{2003BattyeWeller}; \cite{2004MajumdarMohr}). Their redshift mass
detection limits rely only on simple noise estimates, i.e.
the instrumental noise levels, rather than performing realistic simulations and
applying cluster extraction algorithms.

 Hence, so far there has
not been a study that bases its cosmological parameter constraints forecast on
selection and contamination estimates derived directly from realistic
simulations and SZ cluster extraction algorithms. Therefore, this work attempts
for the first time to place constraints on the basis of a realistic Planck cluster
detection pipeline. Here we use the popular
matched filtering 
 technique 
 to assemble a cluster
candidate catalogue, from which by comparison with the cluster input
population the selection and contamination of the cluster samples are derived. 
 Based on
the cluster catalogue properties (selection and purity), we examine the
constraints which can be placed on cosmological parameters, mainly $\Omega_m$ and $\sigma_8$.

The paper is organised as follows. In section \ref{sec:szetheo} we give
 a brief definition of the SZ effect. 
Besides the general composition of (mock) observations at microwave
frequencies, the theoretical basics and the formalism of the used cluster
 extraction method 
 are
described in section \ref{sec:cldetecm}.
Section \ref{sec:skypplanck} summarises details of our performed mock
simulations and discusses their ingredients. 
The implementation 
 of the cluster detection algorithm 
 is 
 discussed in section 
 \ref{sec:mfmfplanck}. 
 Further, in 
 this
section 
 properties of the obtained catalogue, mainly its 
 completeness and
purity, are investigated. 
 Cosmological constraints obtainable from the detectable cluster abundance based on the
efficiency 
 of the extraction method 
 when applied to Planck data (from the Planck
cluster sample) under different
assumptions 
 of the fiducial cosmology, knowledge about cluster redshifts and the Hubble parameter prior are
presented in section \ref{sec:expeccosconstrplanck}. Finally, we close our discussion and conclude in
section \ref{sec:plmfmfmcmcconcl}.

\section{The Sunyaev-Zel'dovich effect in brief}
\label{sec:szetheo}

The anisotropy in the microwave band caused by the SZ effect can be separated into two
contributions which are distinguished by the origin of energy of the
scattering electrons that is responsible for the shift of photon
frequency. The total distortion due to the SZ effect is given by
\begin{equation}
\frac{\Delta I_{\nu}}{I_0} = g(x) y - h(x)\beta \tau \,,
\label{equ:sz}
\end{equation}
where $x=h\nu/k_{{\rm B}}T_0$ with $T_{0}=2.725 \, {\rm K}$
(\cite{1999Matheretal}) and $I_0 =2k_{{\rm B}}^3T_0^3/h^2c^2$.
The first term in equation (\ref{equ:sz}) is the so called thermal SZ
effect due to the thermal motion of electrons of the intra-cluster
gas. The thermal SZ effect has a spectral shape given by
\begin{equation}
g(x) = {x^4 e^{x} \over (e^{x}-1)^2} \, \left[ x {e^{x}+1 \over
e^{x}-1} - 4 \right],
\label{equ:tszspec}
\end{equation}
and a frequency independent magnitude, the Comptonization parameter,
\begin{equation}
y = {k_{{\rm B}} \sigma_{\rm T} \over m_e c^2} \, \int \, n_e T_e \, dl \,.
\label{equ:tszy}
\end{equation}
In hot clusters ($T_e > 5$keV) the relativistic electrons present
slightly modify the spectral shape of the thermal SZ effect (\cite{1998ChallinorLasenby}). This resulting relativistic correction has not been
taken into account in this work, since its effect on
the results presented is negligible. The detectability of the effect
from thermal (e.g. \cite{1998Pointecouteau}) and non-thermal (e.g. \cite{2004EnsslinHansen}) relativistic electrons has been
estimated elsewhere. It is still a matter of debate if Planck will be able to
detect relativistic SZ contributions. The spectral shape of the second contribution in
equation (\ref{equ:sz}), the kinematic SZ effect, is given by
\begin{equation}
h(x) = {x^4 e^{x} \over (e^{x}-1)^2},
\label{equ:kszspec}
\end{equation}
and its magnitude, $\beta = v_{\rm pec}/c$, depends on the uniform
peculiar line-of-sight bulk motion of the cluster's electron plasma,
$v_{\rm pec}$.
\begin{equation}
\tau = \sigma_T \, \int \, n_e \, dl \,,
\label{equ:thoptdepth}
\end{equation}
is the Thomson optical depth. In the case the cluster can be assumed
to be isothermal the Comptonization parameter can be expressed by
\begin{equation}
y = \left( {k_{{\rm B}} T_e \over m_e c^2} \right) \tau.
\label{equ:tsziso}
\end{equation}
In this paper we concentrate on the thermal SZ effect and treat the
kinematic merely as a contaminant to the thermal SZ.

\section{Cluster detection method}
\label{sec:cldetecm}
 
Before describing the method 
 utilised, a brief schematic overview of the nature of
SZ cluster survey observations and known contributing CMB components is given. Based on these considerations an assessment can be made
of the requirements that separation techniques have to
satisfy. More detailed discussions about which components are of
importance to microwave observations at the Planck
observing frequencies are presented in section \ref{sec:skypplanck}.

\subsection{Mock observations}
\label{sec:datavec}

The various components contributing to observations at the Planck frequencies
have either different physical processes and/or different sources as origin. Their
contributions therefore vary with frequency and scale. 
The SZ effect  - the component of interest - as described in the previous
section is due to inverse Compton
scattering of CMB photons off electrons inside galaxy clusters, which are localised extended objects. In microwave observations the
average cluster appears as a source with an extension of the order of a few
arcminutes ignoring instrument dependent beam convolution. Several other components of different nature contribute to SZ observations as a background or foreground. Here, we briefly mention the most important
 ones. 
 First of all,
there is the primordial CMB component. According to standard
inflationary theories, which are in good agreement with constraints placed by
recent observations, this component is a homogeneous random Gaussian field  entirely described by its power spectrum.
Cumulatively, field 
 point sources contribute in an isotropic manner
to SZ observations in the radio and far-infra-red wavelength regime.
 Furthermore, in the Galactic plane
dust, free-free and synchrotron emission from the Milky Way, our Galaxy,
also represent within certain wavelength regimes an important source
of confusion. 
The components mentioned so far are all of cosmological or 
astrophysical nature. The spatial resolution of the observations is limited by the
instrument design and the resulting instrumental beam. 
A component of a different kind, which unavoidably corrupts the observed
data is the instrumental noise. Hence, generally, a SZ observation at a single
frequency $\nu$ can be modelled by:
\begin{equation}
d_{\nu}({\bf x}) = \sum_{i} s_{\nu\,i}({\bf x}) + n_{\nu}({\bf x}),
\label{equ:singlefreqobs}
\end{equation}
where $s_{\nu\,i}({\bf x})$ is the contribution at position
${\bf x}$ of the $i$th cluster and $n_{\nu}({\bf x})$ gives the cumulative
`noise' contribution (including all other components) to the data
$d_{\nu}$ at ${\bf x}$. For a single frequency survey $d_{\nu}$ is a scalar
field. In the following, $s_{\nu\,i}({\bf x})$ refers to the thermal SZ
contribution of the cluster, whereas other SZ components, if present are regarded as noise.
Even though point sources are localised objects, in the following their
collective contribution is
regarded as a single diffuse noise component.

Building on equation (\ref{equ:singlefreqobs}) a multi-frequency observation is described by:
\begin{equation}
{\bf d}({\bf x}) = \sum_{i} {\bf s}_{i}({\bf x}) + {\bf n}({\bf x}),
\end{equation}
where ${\bf d}^t = (d_{\nu_1}, d_{\nu_2}, ..., d_{\nu_n})$, ${\bf s}_i^t = (s_{{\nu_1}\,i},
s_{{\nu_2}\,i}, ..., s_{{\nu_n}\,i})$ and ${\bf n}^t = (n_{\nu_1}, n_{\nu_2}, ...,
n_{\nu_n})$ are transposed column vectors of the data, the $i$th cluster SZ signal and the
noise. Their components are the particular values and 
contributions respectively at observing frequencies $\nu_1, \nu_2, ..., \nu_n$.

\subsection{Matched filtering}
\label{sec:mf}

This section discusses the matched filtering technique, which utilises spatial
as well as spectral information to detect SZ decrements (increments) of galaxy
clusters. The literature also refers to it as optimal filter (e.g. \cite{1996HaehneltTegmark}). In which way the matched filter is an {\em optimal} one
is discussed later in this section. The matched filter is a template
cluster extraction method.  A common cluster template is, for
example, the spherically symmetric $\beta$-profile (see e.g. \cite{1976CavaliereFusco-Femiano}). In this
case the observed SZ signal for the $i$th cluster at position ${\bf x}={\bf 0}$ at frequency $\nu_j$ is given by:
\begin{eqnarray}
s_{\nu_j\,i}({\bf x}) & = & \int B_{\nu_j}({\bf x}-{\bf x}') f_{\nu_j} A_{\nu_{\rm
 ref}\,i} \tau({\bf x}') d^2{\bf x}' \nonumber \\
 & = & \int B_{\nu_j}({\bf x}-{\bf x}') f_{\nu_j} A_{\nu_{\rm ref}\,i} [1 +
 (|{\bf x}'|/r_{c\,i})^2]^{\frac{1}{2}-\frac{3}{2}\beta} d^2{\bf x}',
\end{eqnarray}
where $A_{\nu_{\rm ref}\,i}$ is the amplitude of the $i$th cluster at the reference frequency
$\nu_{\rm ref}$, $r_{c\,i}$ determines the spatial cluster scale (cluster core
radius), $B_{\nu_j}$ is the instrumental beam at frequency $\nu_j$ and
$f_{\nu_j}$ the frequency conversion factor ($f_{\nu_{\rm ref}} = 1$). $\tau$
represents the spatial template normalized to unit amplitude. 

In constructing the matched filter for a given profile, the noise $n_{\nu_i}({\bf x})$ is assumed to be homogeneous with average value $\langle n_{\nu_i}({\bf x})\rangle = 0$ and cross-power spectrum $P_{{\nu_i} {\nu_j}}({\bf k})$ defined by:
\begin{equation}
\langle \tilde{n}_{\nu_i}({\bf k})\tilde{n}^*_{\nu_j}({\bf k}')\rangle = 
P_{\nu_i\nu_j}({\bf k}) \, \delta_{\rm D} ({\bf k}' - {\bf k}),
\end{equation}
where $\tilde{n}_{\nu_i}({\bf k})$ is the Fourier transform of the noise
$\tilde{n}_{\nu_i}({\bf x})$, $\tilde{n}^*_{\nu_i}({\bf k}')$ denotes its complex conjungate and
$\delta_{\rm D}$ is the Dirac delta function.
The homogeneity of the noise ensures that its statistical properties are
independent of position. Cosmic backgrounds such as the primordial CMB and
point sources, as well as the instrumental noise, meet this requirement
of statistical homogeneity. Globally, Galactic components, such as Galactic
dust emission, are not homogeneous. However, on the scale of clusters ($\sim$ several arcmin$^2$)
homogeneity of these components is a reasonable assumption. Our approach to
estimating the background noise cross-power spectrum is discussed in section \ref{sec:skypplanck}.

Moreover, Galactic and point source contributions are caused by emission
 processes. Therefore, they (always) cause physically an increment in the observed
 temperature. This violates the requirement of zero mean and leads to a
 biasing of the recovered signal amplitude of the $i$th cluster, $A_{\nu_{\rm
 ref}\,i}$. 
 Assuming a central limit 
 and taking off the zero
 Fourier mode\footnote{Here it is assumed that the emission components' pixel
 temperature values scatter 
 roughly symmetrically about their mean.
}, the
 map is modelled in simulations to have a zero mean. 
This is a fairly safe
 modelling approach of CMB sky observations since the monopole on the targeted
 sky patch is usually unobserved by common
 instrumental designs.
 The required $\langle n_{\nu_i}({\bf x})\rangle = 0$ can thus be satisfied.
The fact that the zero Fourier mode (monopole) is unobserved has a negligible effect
on measured cluster fluxes.

Given the assumptions of zero mean and spatial homogeneity the optimal matched filter
is then derived as follows.  If the sky is observed at $n_f$ frequencies, the most general linear estimator
of the amplitude $A_{\nu_{\rm ref}i}({\bf x})$ of cluster $i$ at position
${\bf x}$ is given by:
\begin{eqnarray}
A^{\rm est}_{\nu_{\rm ref}\,i}({\bf x}) & = & \int \Psi({\bf x}-{\bf x}')\cdot {\bf d}({\bf x}')\,\mbox{d}^2{\bf x}' \nonumber \\
& = & \sum^{n_f}_{j=1} \left(\int \psi_{\nu_j}({\bf x}-{\bf x}')\,
d_{\nu_j}({\bf x}')\mbox{d}^2{\bf x}'\right),
\label{equ:est}
\end{eqnarray}
where $\Psi$ and ${\bf d}$ are column vectors each containing $n_f$
components. 
 The components of $\Psi^t = (\psi_{\nu_1}, \psi_{\nu_2}, ...,
\psi_{\nu_{n_f}})$ are frequency dependent weight
functions. $A^{\rm est}_{\nu_{\rm ref}\,i}({\bf x})$ represents the estimate
of the cluster SZ signal
amplitude. This convolution (equation \ref{equ:est}) can be written in Fourier
space conveniently as a product:
\begin{equation} 
A^{\rm est}_{\nu_{\rm ref}\,i}({\bf x}) = \sum^{n_f}_{j=1} \left(\int 
\tilde{d}_{\nu_j}({\bf k})\tilde{{\psi}}_{\nu_j}({\bf k}) e^{- i{\bf k}\cdot {\bf x}} \mbox{d}^2{\bf k}\right),
\end{equation}
where $\tilde{d}_{\nu_j}$ and $\tilde{\Psi}_{\nu_j}$ denote the Fourier
transform of the data and filter at frequency $\nu_j$.

In the case of the matched filter one requires the weight
function vector $\Psi$ (the filter) to satisfy the following criteria:

\begin{enumerate}
\item{The quantity $A^{\rm est}_{\nu_{\rm ref}}$ is an {\em unbiased} estimator 
of the SZ amplitude of the cluster. Thus $\langle A^{\rm est}_{\nu_{\rm ref}}\rangle = A_{\nu_{\rm ref}}$ is required.}
\item{The variance of the noise of the estimator, $\sigma^2$, is minimized by the filter, which ensures that $A^{\rm est}_{\nu_{\rm ref}}$ is an {\em efficient} estimator.}
\end{enumerate}

The requirement of being an unbiased estimator fixes the normalisation of the
filter by:
\begin{equation} 
\langle A^{\rm est}_{\nu_{\rm ref}}\rangle - A_{\nu_{\rm ref}} = \int
\Psi({\bf x}) \cdot {\bf s}({\bf x})\mbox{d}^2{\bf x} - A_{\nu_{\rm ref}} = 0,
\end{equation}
where the brakets $\langle\rangle$ denote an average estimate for a specified spatial
cluster template centred at the origin (${\bf x}={\bf 0}$) obtained over many noise realisations.
The filter shape is determined by demanding the variance of the estimate,
\begin{eqnarray} 
\sigma^2 & = & \langle (A^{\rm est}_{\nu_{\rm ref}})^2 \rangle - \langle A^{\rm est}_{\nu_{\rm ref}} \rangle^2,
\end{eqnarray}
to be minimal. This minimization of $\sigma^2$ ensures that the shape of the
filter is {\em optimally} chosen to be maximally sensitive to modes at which
the cluster signal exceeds the noise. The multi-frequency filter satisfying these
conditions is given in Fourier space by the matrix equation (see also \cite{1996HaehneltTegmark};
\cite{2002Herranzetal})
\footnote{Due to the fact that clusters have on average an
  angular scale of a few arcminutes, the flat sky approximation and therefore working in
  Fourier instead of spherical harmonic space is an adequate approximation.}:
\begin{eqnarray} 
\tilde{\Psi}({\bf k}) & = & \alpha
{\bf P}^{-1}({\bf k})\tilde{{\bf F}}({\bf k}), \\
{\alpha} & = & \left(\int {\tilde{{\bf F}}}^t({\bf k}) \, {\bf P}^{-1}({\bf k}) \,
{\tilde{{\bf F}}}({\bf k}) \, d^2{\bf k} \right)^{-1} \nonumber \\
 & = & {\sigma}^2,
\label{equ:mfmf}
\end{eqnarray}
where $\tilde{{\bf F}}$ is a column vector described by $\tilde{{\bf F}}^t =
(f_{\nu_1} \tilde{B}_{\nu_1} \tilde{\tau}, f_{\nu_2} \tilde{B}_{\nu_2} \tilde{\tau}, ..., f_{\nu_n} \tilde{B}_{\nu_n} \tilde{\tau})$ and ${\bf P}^{-1}$ is the
inverse of the noise cross power spectrum matrix with components
$P_{\nu_i\nu_j}({\bf k})$. Equation \ref{equ:mfmf} gives the so-called 
multi-frequency matched filter (MFMF), which is an extension of the
single-frequency matched filter (SFMF) derived in \cite{1996HaehneltTegmark}. 

\section{Microwave sky simulations}
\label{sec:skypplanck}

\begin{figure*}
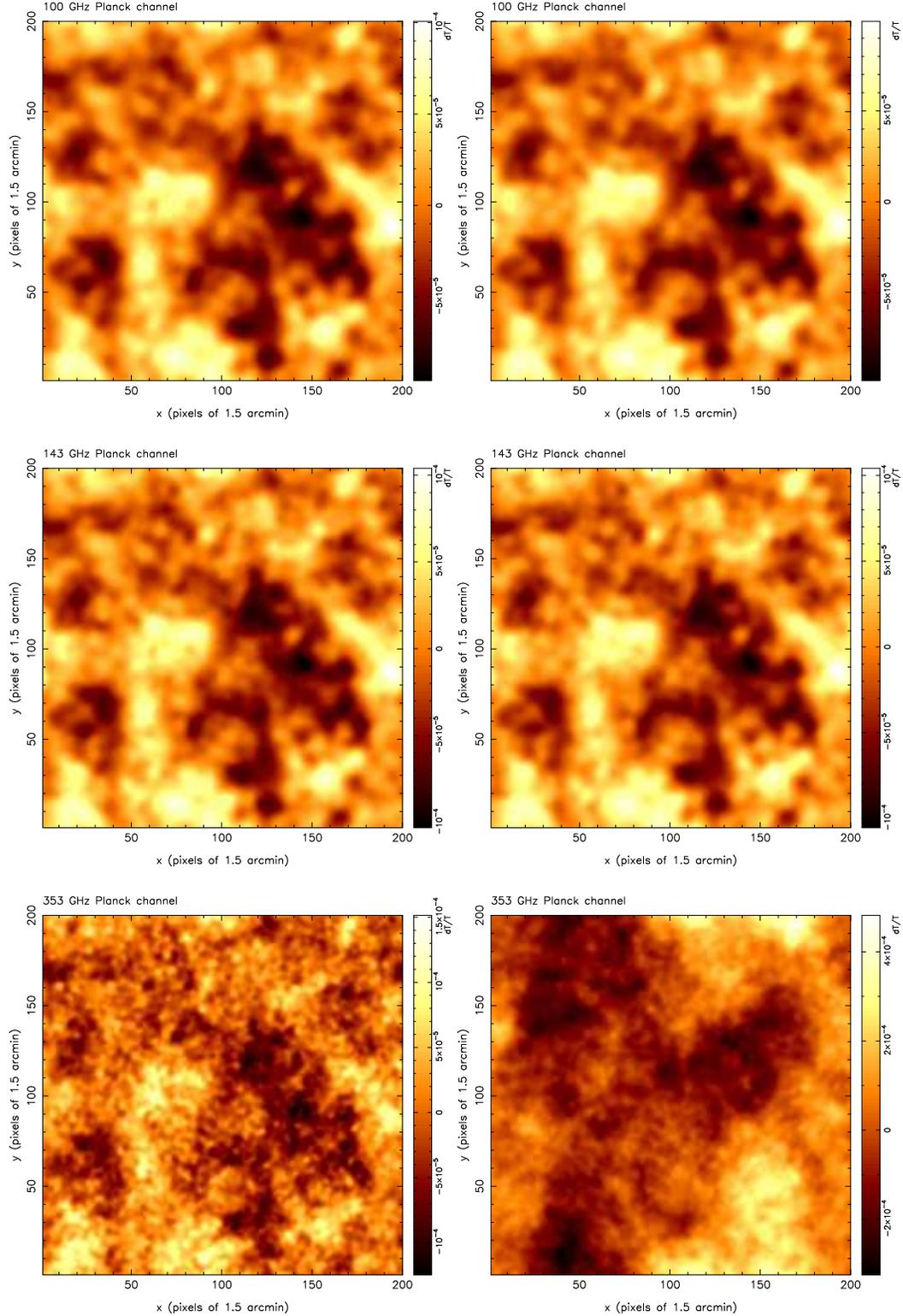

\begin{center}
\epsfig{file=figs/mfmf/data_planck_2_01.ps,width=6.5cm,angle=-90,clip=}
\epsfig{file=figs/mfmf/data_planck_1_01.ps,width=6.5cm,angle=-90,clip=}\\
\bigskip
\epsfig{file=figs/mfmf/data_planck_2_02.ps,width=6.5cm,angle=-90,clip=}
\epsfig{file=figs/mfmf/data_planck_1_02.ps,width=6.5cm,angle=-90,clip=}\\
\bigskip
\epsfig{file=figs/mfmf/data_planck_2_03.ps,width=6.5cm,angle=-90,clip=}
\epsfig{file=figs/mfmf/data_planck_1_03.ps,width=6.5cm,angle=-90,clip=}
\caption{$5\times5$ deg$^2$ realisations of Planck observations at the three
  observing frequencies of the satellite which are most important for cluster
  detection via the thermal SZ effect. The upper panels show the observed
  patches at 100 GHz, the mid panels show the observed patches at 143 GHz and the
  bottom panels show the observed patches at 353 GHz. The left column shows a
  patch lying outside the Galactic plane and the right column a patch within
  the Galactic plane. The same pixelisation scheme and primordial CMB
  realisation have been adopted for each channel map.}\label{fig:planckclskypatch}
\end{center}
\end{figure*}

In succeeding sections, extraction algorithms are applied to simulated Planck
observations of $5\times 5$deg$^{2}$ sky patches. In particular, data is
simulated for the Planck High Frequency Instrument (HFI) channels at 100 GHz,
143 GHz, 217 GHz, 353 GHz, 545 GHz and 857 GHz. Due to their resolution
and/or the ratio of the SZ signal in comparison to amplitudes of other fore-/backgrounds within the
channel bandwidths, three of these HFI frequency channels - namely the 100
GHz, 143 GHz and 353 GHz channel - are the most useful ones of
Planck for galaxy cluster detection via the SZ effect.\footnote{Note that
 Planck channels which have not been taken into account in this analysis provide extra
 information about the SZ signal and contaminants. Adding these channels in
 the analysis hence causes a slight increase in the number of detected
  clusters and decreases the number of false detections. Trial runs suggest
  that this affects the total cluster number count by less than 10 per cent. The
  obtained completeness estimate is thus a conservative lower one. We
  restrict our analysis here to the HFI channels including the three most important frequencies for SZ
  cluster detections to keep computational cost down. For the purpose of
  estimating cosmological constraints derivable from the Planck cluster
  sample, effects on the cluster number count of this order are of minor relevance. Moreover, other effects
  such as the
  uncertainty of the mass function are of similar order of magnitude.} Simulated sample
patches outside and within the Galactic plane observed at the frequencies of these three channels are shown in Figure \ref{fig:planckclskypatch} to give an
impression of the expected quality of the Planck mission data. For the purpose
of estimating the performance of the extraction techniques when applied to Planck data,
patches residing within the Galactic plane and outside of it
have been simulated in the right proportion. In the
case that observed patches lie within the Galactic plane, the simulations have
to include modelling of the Galactic dust, synchrotron and free-free
emision besides the primordial CMB,
the SZ effects, extragalactic point sources and instrumental effects. Actually,
Galactic dust is by far the most dominant Galactic component at the
frequencies of interest. 
 Realisations of the Galactic components, the primordial
CMB and the SZ effects have been obtained similar to the ones described in GKH05.
However, the dust modelling uses this time a two temperature model. Moreover,
the anisotropic nature of the Planck instrumental noise on the sky due to the
scan pattern of the satellite is taken into account.
The extragalactic point source population has been modelled in a different way
 than previously.
 Instead of utilising
the theoretical model of \cite{1998Toffolattietal} as in GKH05, which has been found to
match the WMAP point source detections within a factor of two at frequencies
below 100 GHz, a phenomenological approach has been taken this time to obtain
number counts of the radio and far infra-red/sub-mm point sources at the Planck
observing frequencies. Extrapolations of WMAP
 data (\cite{2003Bennettetal}) suggest that the contamination due to radio
 sources is marginal above 100 GHz. While radio point sources dominate below 100 GHz, at
 the Planck channel frequencies of interest ($\nu \geq 100$ GHz) the confusion caused by
 dusty luminous infra-red sources is most important. 
The extragalactic far infra-red/sub-mm (IR/SM) source count modelling
 performed here is
 based on the 350 GHz observations of the Submillimetre Common User Bolometer
 Array (SCUBA; \cite{1999Hollandetal}) mounted on the James Clerk Maxwell
 Telescope (until 2003). There have been several deep observations made with SCUBA
 (\cite{1997Smailetal}; \cite{1998Bargeretal}; \cite{1998Hollandetal}; \cite{1998Hughesetal}; \cite{1999Ealesetal}) from
 which source counts have been obtained. 
 SCUBA blank field counts of the IR/SM source population on larger
 fields have been obtained by \cite{2002Scottetal},
 \cite{2003Borysetal} and \cite{2006Scottetal}. From this work one can derive
 phenomenological fitting formula to the source counts at 350 GHz.
The extrapolation of these source counts to lower frequencies ($100 \leq \nu
 \lesssim 300$ GHz), however, involves substantial uncertainties since
 the spectral behaviour of the sources is not (very) well known and may change
 significantly from one point source to another. Furthermore,
 due to a general lack of observations in the frequency regime, knowledge of the emission of extragalactic point sources at
 intermediate HFI Planck channels is relatively poor and one has to rely on
 extrapolations based on assumptions about the source spectra. In the (rather
 unlikely) case
 of precise knowledge of the spectral behaviour, one could measure the source
 flux at higher frequencies ($\nu \geq 350$ GHz) at which these sources
 dominate and subtract off appropriate flux levels at the frequencies of
 interest. Here, we
 take a mean spectral index of $\alpha=2.6$ and assume a rms scatter of
 $\sigma_{\alpha}=0.3$ around this mean for individual galaxies to do the
 spectral extrapolation. 
We further assume that IR/SM sources relevant to Planck observations are
spatially uncorrelated with clusters detectable by Planck. Given that luminous
dusty galaxies are expected to be at high redshifts ($z>1$) this should
be a reasonable assumption.

Flat sky patches can be obtained from full sky Planck data by stereographic 
projection of reasonably sized regions of the
sphere onto planes tangential to the sphere at the centres of the patches.\footnote{Note that the patch size should not exceed $15$
  degrees at most to avoid significant structural deformations.}
 Figure
\ref{fig:planckclskypatch} shows simulated observations of example patches at
three considered Planck channels. Even though the patch shown in the left
column lies outside the
Galactic plane (low dust region) the identification of galaxy clusters by eye
in the map is impossible.

This approach of splitting up the sky into patches is in the case of optimal
  filtering the preferred one, 
 since background noise levels vary significantly
over the full sky. For the matched filter method, it is therefore necessary to have some
  estimate of the local background noise power spectrum. This knowledge has been
assumed to be available 
(to some realistic degree). For example, it can be gained
from the HSMEM separation which has the ability to recover several physical
  components at ones, each of which is spatially distinct and has a different
  frequency dependence. 
However, as a zeroth order estimate of the noise power spectrum one may also
 take the power spectrum of the sky patch under consideration
 and then
 iterate until the extracted cluster number and the power spectrum estimate of
  the noise converge. Explicitly, this is done by removing in each iteration
  step the number of clusters detected above a threshold (e.g. above a
  signal-to-noise threshold of 3$\sigma$) from the map and taking the residual
  map as the new background noise estimate.
 If background noise estimates are wrong, the detection
significance (signal-to-noise ratio) returned by the method is systematically
 flawed - hence this is a source of systematic error.

Before applying a MFMF cluster extraction algorithm to the data, one can
 perform a \mbox{(pre-)cleaning} of the channel maps to reduce the level of certain contaminants and
 increase the signal-to-noise ratio of the SZ signal. Point sources, for example, can be removed by a Mexican Hat
 wavelet technique (see e.g. \cite{2001Vielvaetal}). 
 Since the 857 GHz channel is completely and the 545 GHz
 channel mostly dominated by
Galactic dust emission within the plane of our Galaxy after the removal of
 IR/SM point sources, these channels can be used on their own or in combination to remove the dust emission of the Milky Way at lower frequencies. Furthermore, one might lower the level of
 primordial CMB contamination by using the 217 GHz channel and performing
 spatial filtering in spherical harmonic/Fourier space.
\cite{2002Herranzetal} found that if all this cleaning is performed the cluster number
 count obtained by a MFMF method
 above a 3$\sigma$ detection threshold is increased by 7 percent and the false
 detection rate lowered by 12 percent in comparison to the results obtained
 when the MFMF scheme is applied directly to the `raw' data. At higher
signal-to-noise detection thresholds this gain is expected to be even
 less (in the following a $5\sigma$ detection threshold is used). This suggests that even on its own a MFMF method yields robust and
 reliable results. Thus such a \mbox{(pre-)cleaning} step has not been included in
 our analysis pipeline.

\section{MFMF cluster extraction}
\label{sec:mfmfplanck}

In the following we apply multi-frequency matched filters to the sky
observation simulations described above. The filters are constructed according
to the instructions given in section \ref{sec:mf}.
A cluster candidate catalogue is compiled by using a cluster extraction
algorithm based on multi-frequency matched filtering consisting of a number of
steps explained in this section.

\subsection{Extracting thermal SZ cluster signals}
\label{sec:tszmfmfplanckalg}

Apart from the
initial convolution of the multi-frequency data with the diverse filter
kernels to generate a multi-dimensional detection likelihood space, the
extraction algorithm consists of several steps which are iteratively repeated
until all candidates with detection significances above the required
threshold are obtained. The algorithm is similar to the one suggested by
\cite{2003SchulzWhite}.

First, the frequency channel maps are convolved with filters whose spatial
scales are gradually varied.
The upper and lower limit of the spatial filter scale are
determined by the expected range of sizes of detectable clusters on the
sky. Besides, performing scale dilation the shape of the cluster template
used to construct the filter kernels
can also be varied. 
 Since the complexity of several steps of the
algorithm scales (linearly) with the diversity of the filter kernels, for a
given patch of fixed pixel resolution the computational cost of the algorithm
depends strongly on the number of distinct filter kernels applied. The optimal
discretisation of the
kernels depends on the data at hand. For example, due to the
 beam sizes of the Planck channels which are rather large in comparison to the
 average cluster size, one does not expect to gain (much) valuable information about the
scales of the unresolved majority of clusters.\footnote{It is referred to a
  cluster as being unresolved in the case that the cluster's entire flux is picked
  up by
  one instrumental beam pointing. The cluster therefore appears as a point
  source in the Planck data.} Thus, a rather coarse filter scaling should be
adequate on scales not resolved by the Planck beams. However, as we discuss 
in section \ref{sec:tszmfmfplanckremarks} otherwise a fine filter scale
discretisation is
 advantageous. In order to construct a detection
likelihood space, the convolved
multi-frequency channel maps are co-added and normalised to unit variance for
each filter kernel as described in section \ref{sec:mf}.

Subsequently, at each iteration the cluster candidate with the highest signal-to-noise
ratio is identified in the unit variance normalised `detection likelihood space'
spanned (in our case) by the position parameters
and the discretized filter scale. A bright cluster has a high
detection significance at various filtering scales at (roughly) the same sky
position. Its photometric 
 parameters are determined on the
basis of the variation of the detection likelihood (peak
height in the normalised likelihood space) with filter scale. For the filter
scale that is closest to the true scale of the cluster, its signal-to-noise ratio should become maximum. The detection is
added to the cluster candidate list and its estimated signal is subtracted from the
cube. This removal takes place directly within the detection likelihood
space. It is realised by subtracting off an amplitude normalised template at
the candidate's sky position. The
template consists of the unit peak (most likely) cluster candidate
shape convolved at each scale by the respective filter kernel. Since the
number of filter scales is limited,\footnote{This causes the best scale
  estimates of the candidates to be discretised as well.} the template can be
\mbox{(pre-)generated} and stored for (further) applications of the
unaltered 
 algorithm. The amplitude to which the template is finally
normalised corresponds to the most likely central cluster candidate SZ
distortion in units of the standard deviation of the co-added filter kernel map.

 Thereafter, the algorithm is reapplied. The procedure commencing
 from the step of locating the most significant detection is repeated in a loop on
the residual detection likelihood maps until no further detection is found above the
chosen signal-to-noise threshold.
Thus the extraction algorithm represents an iterative approach similar to a
CLEANing procedure (see e.g. \cite{1974Hoegbom}; \cite{1980Clark}).
This procedure represents a convenient way to disentangle cluster-cluster
confusion as long as the detection of highest significance coincides with the
brightest remaining cluster candidate on the patch. This is usually
the case. Therefore, by its removal the signal-to-noise ratios of the rest of
the candidate detections become unbiased. However, occasionally due to 
 biasing the signal-to-noise ratio of a candidate is overestimated.
For example, in the case that filtering at various scales indicates that the projected SZ signal of
several clusters might overlap and the single candidate detection significance might be
biased due to the overlap or even the detection might be a false one, one has
to (re-)assure that the detection significances are to the largest possible
extent unbiassed. 
 This can be tested by varying the order of iterative
removal of the candidate detections under consideration. It is the candidate
whose signal-to-noise ratio varies the least which should be
removed first.
Such overlaps sometimes occur by chance due to line-of-sight projection over
a deep light-cone, as well as in supercluster environments in which clusters
are situated close to each other even in redshift space.

Moreover, splitting up the observed patch into seperate regions, in such a way
that none of the detections in one region is affected noticeably
by one of the other regions and
vice versa, speeds up the algorithm slightly by reducing the number of iterations required until
all candidate detections above a specified detection threshold are
found. The number of iterations\footnote{Here we refer
  to an iteration as a step that ends with the listing and (complete) removal of
one or several clusters.} reduces from the number
of all candidates to the highest number of candidates in one of the seperate regions
above the signal to noise threshold. However, the speed up is mostly due to
the minimisation of space in which operations have to be performed. It also
provides a way to parallelise the extraction algorithm when it is performed on
large datasets.
The angular clustering of galaxy clusters and the largest scale covered by the set
of filter kernels place a lower limit on the patch (region) size.
The size has always to be chosen large enough so that all clusters affecting each
other are located in the same patch (region).

After completion of the algorithm, a cluster candidate catalogue containing
the candidate's position on the sky, scale, central amplitude, flux,
morphological parameters, such as the (asymptotic) slope of the profile,
ellipticity and inclination angle,\footnote{For the sake of minimising
  computational cost we did not vary morphological parameters of the profile
  to create filter kernels. We rather assumed the cluster profile which has
  been used to simulate the SZ sky to be known. However, this makes no
  difference for unresolved clusters. In the case a cluster is resolved
  adequate `morphological' variation of the filter kernel should
in the limit return results as presented here.} (etc.) is at hand.

\subsection{Cluster catalogue properties}
\label{sec:tszmfmfplanckcat}

\begin{figure}
\begin{center}
\epsfig{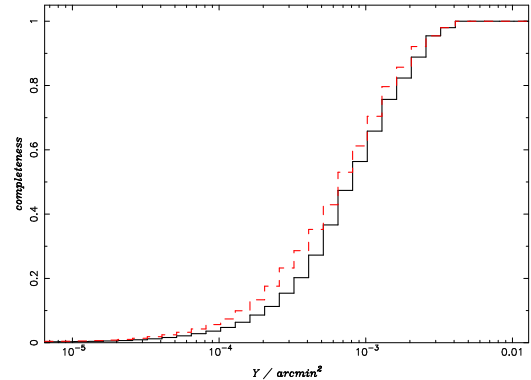}\\
\caption{The integrated completeness of the cluster catalogue obtainable from the Planck
  survey using the MFMF algorithm. Besides matching candidates with single clusters, candidate
  detections have also been allowed to have several clusters associated with
  them due to projection along the line-of-sight onto the sky. Single cluster
  matching is given by the solid line, while multi cluster matching is shown
  by the dashed line in each case. The integrated completeness is defined as
  $\mathscr{C}(Y_{\rm cl} \geq Y,z \geq 0) = N_{\rm det}(Y_{\rm cl} \geq Y,z
  \geq 0)/N_{\rm exp}(Y_{\rm cl} \geq Y,z \geq 0)$, where $N_{\rm det}(Y_{\rm
  cl} \geq Y,z \geq 0)$ is the recovered number of clusters above a flux $Y$
  and $N_{\rm exp}(Y_{\rm cl} \geq Y,z \geq 0)$ is the expected cluster number
  above $Y$ over the whole sky and redshift range.}\label{fig:planckszcatcompl}
\end{center}
\end{figure}

After applying the MFMF algorithm to a representive number of sky patches of
$5\times 5$ deg$^2$ whose noise realisations (instrumental noise levels,
Galactic foregrounds) are varied according to the proportion they occupy
on the full sky, the properties of the obtained cluster candidate catalogue
can be evaluated. For this purpose the candidates have to be associated with
real clusters.

\subsubsection{Matching up candidates with clusters}

Based on the extracted cluster candidate list a matching is performed
between candidates and clusters of the input cluster catalogue.
Similar as in GKH05, a lower matching flux limit of $Y=3\times 10^{-4}$ is
adopted to avoid dubious associations occuring just by chance. This flux
threshold corresponds approximately to the analytically derived $3\sigma$ point source
sensitivity of Planck (see \cite{1997Haehnelt}; \cite{2001Bartelmann}). A
candidate is successfully matched up with a real cluster if the distance
between their central positions does not exceed a predefined matching
length. The association of candidates with clusters is a crucial point
in the evaluation of the performance of the extraction method since it affects
directly the completeness and contamination estimates. For example, in GKH05 a matching length of just one pixel of $\sim 2$ arcmin was assumed
for assigning clusters to candidates recovered from simulated Planck
data. Such a conservative matching length leads to firm lower limits
of the completeness and purity of the recovered candidate sample since every candidate that
is not associated with a cluster due to the small matching length is
considered to be a false detection.
Henceforth, the extent of the matching length is chosen in a way to
account for the fact that cluster candidate positions derived from Planck data
are not highly accurate. Positional uncertainties arise mainly due to
limited instrumental resolution and noise variations on the resolution scale
since cluster core regions even of resolved clusters commonly fit into the
Planck instrumental beams. Therefore, we accept a match if the positional
deviation (of the pixel centres) does not exceed a matching length of
$\sqrt{2}\times {\rm FWHM}$ of the instrumental beam. As the resolution
differs with channel, the average FWHM of the beams is taken
for constructing the matching distance. Moreover, the pixelisation of the maps
is chosen fine enough to ensure that pixelisation effects do not play a (major) role.

In the case that a candidate can be associated with several clusters, it is
matched up with the cluster that yields the best flux match. In the reciprocal
case of several candidates being matched to one real cluster we proceed in the
same way by keeping the best flux match. However, in the
following we also consider that multiple clusters are associated with one
candidate detection and vice versa. Due to the angular clustering of sources and the fairly large
Planck beams, a disentanglement of contributing sources is sometimes
impossible and a candidate's flux estimate can correspond to the sum of
several unresolved sources (of even similar flux). In rare cases resolved low
redshift clusters are detected as multiple candidates due to noise and
background SZ variations on scales smaller than the cluster extent. Then the
candidate which fits best the photometry of the real cluster is assigned to it. All the others are
conservatively regarded as false detections. Liberally, they are simply disregarded in the case they fall within the matching region of the accepted candidate.
As already mentioned above, if a candidate is not matched up with any real
cluster above the flux limit, it is regarded as a false detection.

\subsubsection{Catalogue completeness and purity}

Important measures of the quality of a recovered cluster catalogue are its
completeness and purity. These also set benchmarks for the evaluation of how
useful the catalogue can be for cosmological purposes. For example, a low
completeness results in large uncertainty of the total number of clusters on an
observed patch. A catalogue becomes useless if besides a low completeness it
possesses as well a low purity. The few positive detections are then diluted by false ones
and the catalogue can neither be used to make predictions about the cluster
ensemble nor serve as a basis for follow-up observations to learn more about individual objects.
If one does not want to rely heavily
on observations at other wavelengths (e.g. optical and X-ray cluster
observations), which are possibly spoiled themselves, the only way to
determine these measures, the completeness and purity, is to perform realistic simulations whose
ingredients, such as the underlying cluster sample, are known.

Figure \ref{fig:planckszcatcompl} shows the completeness of the catalogue
extracted by the MFMF algorithm for candidates detected above a
signal-to-noise threshold of 5. Above $Y \sim 10^{-3}$ arcmin$^2$ the vast majority of
clusters is recovered. At $Y \approx 10^{-3}$ arcmin$^2$ the sample is
 still more than 50 percent
complete. Below this flux regime the completeness of the sample falls steeply.\footnote{Note that
  for visualisation purposes the flux ($x$-axis) is plotted logarithmically in
Figure \ref{fig:planckszcatcompl}. This and the use of the integrated completeness weaken the impression of steepness of
the 
 curve. The remaining marginal completeness at low fluxes is entirely due to
 cluster detections at higher fluxes.} 
 The 
 fraction of clusters with a 
 real flux $Y<10^{-3}$ arcmin$^2$
that is matched up with a candidate of 5$\sigma$ detection significance is low. As
expected, the completeness is increased in the case that multiple clusters are
permitted to be assigned to one candidate (see dashed line in Figure \ref{fig:planckszcatcompl}). At high cluster fluxes this is mainly
due to halo clustering in redshift space (supercluster environment). Due to
the low surface density of (massive) high flux clusters on the sky, cluster overlap is
unlikely to happen just at random (the number density of clusters above the chosen
matching flux limit of $Y=3\times 10^{-4}$ arcmin$^2$ is well below one cluster
per square degree for the assumed cosmological models). Note as well that massive clusters are more
strongly clustered than low mass ones. 
However, the probability
of cluster-cluster confusion to occur by chance due to
projection along the line-of-sight increases rapidly with decreasing flux.
The shown completeness estimate gives an
  average as expected for a full-sky survey. Note that spatially the completeness
varies strongly, depending on the instrumental noise and the Galactic foregrounds.

Further, it turns out that the recovered candidate catalogue is of high purity. The rate of
false detections contaminating the sample of candidates with detections of a
signal-to-noise ratio of $\geq 5$ is of the order of one percent.
Note that due to a limited
number of extracted cluster candidates there is always a sample variance error on
the estimated purity as well as on the completeness estimate. However, by
simulating and analysing an appropriate number of patches which ensures a high number of
extracted candidates, the error on the estimates is minimised and a contamination of the
candidate sample above $\sim 3$ percent can be excluded at high significance. 
On the basis of the completeness of this fairly pure recovered cluster sample, cosmological
parameter constraints are derived in section \ref{sec:expeccosconstrplanck}. In the case the detection
threshold is lowered, as expected, the completeness as well as the contamination increase.

\subsection{Some remarks}
\label{sec:tszmfmfplanckremarks}

Matched filtering is a
template-based object detection approach. To extract the thermal SZ effect,
templates empirically derived from fits to observations (e.g. the
$\beta$-profile: \cite{1972King} and \cite{1978CavaliereFusco-Femiano}) and/or by
hydrostatic theoretical considerations (e.g. \cite{2002KomatsuSeljak} and \cite{2002CooraySheth}) are
utilised. The filter spatial scale is
varied by changing the characteristic radius of the cluster template (e.g. the
core radius of a $\beta$-profile). Furthermore, one may also parametrise the
template shape (e.g. variation of $\beta$ in the case of the
$\beta$-profile). The universality of the
template is essential to guaranty a photometrically unbiased cluster candidate sample
whose signal-to-noise ratio distribution is not systematically skewed. A universal
template represents thus an average scalable shape of a cluster SZ imprint in
the CMB. Due to different cluster environments and morphologies, the imprints
caused by single clusters `scatter' around the average one.

In the case of Planck whose highest resolution of a frequency channel
map is 5 arcmin (FWHM), exact knowledge of the cluster SZ template is of less importance than it is the case for
high-resolution observations (such as AMI observations), since the convolving instrumental beam erases much of
the information on cluster (sub-)structure. Hence, 
 good knowledge of the beam shape
is important in the case of Planck. In previous work (\cite{2003SchulzWhite}) concerning a matched
filter cluster recovery from Planck observations, clusters have been
regarded as point sources and the beam shape has been used as template.
However, even for `low resolution' Planck data
varying template parameters (i.e. building a discretised template library) is advantageous in comparison to a single filter
kernel 
 and
leads to a larger number of extracted clusters. 
 Nevertheless, the finer the discretisation of the template
parameters for constructing matched filtering kernels is, the higher
the number of reliably extracted candidates becomes. 
The computation time needed by the algorithm is raised approximately
linearly with filter kernel variety. At some point, however, the
increase in consumed computing power
outweighs the gain in recovered clusters.
Therfore, there is always a trade-off between
computational cost and maximising the number of reliably extracted candidates. 
Our
implementation of the MFMF is tuned to be most efficient for Planck data with
regard to computational cost and cluster extraction.

The candidate sample (flux) completeness above a chosen flux threshold (see
Figure \ref{fig:planckszcatcompl}) has been derived on the
basis of the concordance model and Planck's instrumental properties. The
completeness is no doubt very sensitive to the instrumental design defining
the instrumental noise level, resolution and number of frequency
channels. Planck's instrumental properties can be expected to be well known.
On the contrary, the noise level due to cosmological contaminants
depends like the cluster abundance on the cosmological model of our Universe.
While the dependence of the primordial CMB on cosmological parameters is well
understood, there is still little known about the cosmological dependence of
number counts, auto-/cross-correlations and evolutions of point source
populations at Planck's observing frequencies.
Hence, we modelled the point source count empirically on the basis of recent
observations. The primordial CMB power spectrum in our simulations has been
chosen so that its shape agrees with
  observational findings from WMAP, VSA and other experiments.
Furthermore, we studied the impact of variations of the cluster (surface)
number density with cosmology on the completeness.
Our testing shows that the completeness is fairly insensitive to such
variations at least for reasonable changes of cosmological
parameters. This insensitivity can be explained by the in comparison to
angular sizes of SZ sources fairly large Planck channel beams which smooth
fluctuations in the SZ background. In the case that any significant
discrepancies between our assumptions and future observations arise, the
algorithms can be rerun on updated simulations to adjust the completeness estimates.   
Moreover, the absolute
number of clusters which will be detectable by Planck depends on several factors. 
Apart from the cosmology of the Universe which heavily influences cluster
detection numbers, also cluster physics plays an important role (see
e.g. \cite{2004daSilvaetal}).

\section{Expected constraints on cosmology from the Planck SZ survey}
\label{sec:expeccosconstrplanck}

It is well known and one of the major motivations of blank field SZ cluster
surveys that the cluster abundance and redshift distribution is sensitive to
cosmological parameters. 
 On the
basis of the `blind' cluster extraction algorithm pursued above and its 
 estimated cluster flux selection (the sample completeness at flux $Y$ and
redshift $z$, $\mathscr{C}(Y,z) = N_{\rm det}(Y,z)/N_{\rm exp}(Y,z)$) 
 and
purity, in the following the constraints one can obtain
from a Planck SZ survey on cosmological parameters are investigated. 

\subsection{Analysis methodology}

The cluster flux dependent selection is found to be approximately
universal at redshifts $z\gtrsim 0.1$ for the implemented SZ extraction
algorithm. The effect of
approximately constant flux sensitivity at redshifts above $z \approx 0.1$ is
essentially due to the vast majority of clusters at these redshifts being
unresolved.
A strong exception to this universality occurs only at very low redshifts. Below
$z\lesssim 0.05$ the limiting flux at which the sample has a
specific constant completeness
increases rapidly to higher fluxes with decreasing redshift. However,
since the affected volume is small (low redshift), the number of clusters missed
is marginal so that the completeness above a limiting flux at redshift $z$
($z\gtrsim 0.1$) does not differ from the redshift integrated one as shown in
Figure \ref{fig:planckszcatcompl} by a large margin. Moreover, all clusters at
such low redshifts ($z\lesssim 0.05$) and with SZ fluxes comparable to those
in the Planck sample should be (easily) detectable by other observational
means. For example, they should have been detected by the ROSAT All Sky Survey (RASS).

For the comparison of the theoretical predictions of the fiducial models
with the ones of other models and subsequently for estimating the constraining
power of the cluster sample at hand on cosmological parameters, we use a MCMC
analysis. Our analysis is based on the Metropolis algorithm
(see \cite{1949MetropolisUlam}) to sample
the (log-)likelihood function over the parameter space. This represents an
efficient way of sampling. For the analysis, our
basic parameter set consists of four cosmological parameters, $\Omega_m$,
$\Omega_\Lambda$, $\sigma_8$ and $h$.
Note that we do not assume that the Universe has a flat geometry, as it has
been often done in recent works of other authors (see
e.g. \cite{2003BattyeWeller} who put their emphasis on constraining the
`nature' of dark energy with SZ cluster surveys). Other parameters which generally can be varied as well
are kept fixed. For example, the spectral index is fixed to $n_s = 1$
(Harrison-Zel'dovich spectrum) and the baryon density $\Omega_b$ is set
to the best fit WMAP value. Furthermore, our presented analysis is restricted
to $\Lambda$CDM models ($w=-1$). The inability of cluster surveys on their own
to constrain the Hubble parameter, causes us to place in the course of our
analysis tight constraints on $h$
obtained by other means, i.e. constraints from the Hubble Space
Telescope Key Project (Gaussian prior: $h=0.7\pm0.08$). Nevertheless, we first
examine the effect of a loose uniform $h$ prior on constraints of other
parameters. The parameter space spanned by the other cosmological parameters
($\Omega_m$, $\Omega_\Lambda$ and $\sigma_8$) is uniformly sampled in our analysis.

Moreover, in the presented work, running a self-calibration analysis, in
which apart from cosmological parameters cluster physics parameters are varied
as well, has not been attempted. The normalisation of the mass-Comptonisation
parameter (mass-observable) relation of
clusters has been assumed to be a priori known and it has been assumed to not
evolve with redshift. However, due to the large
number of clusters, which is of the order of $10^3$ for a detection
significance of $5\sigma$ and probable cosmologies of the Universe, such a
self-calibration analysis is feasable without abandoning completely meaningful
constraints on parameters. Since there is a rather large error on the
reconstructed cluster fluxes, only total cluster numbers on the full sky and
within chosen redshift bins are used here to derive parameter constraints.
Investigating and understanding reconstructed cluster fluxes and deducing a
reliable relation between them and
real cluster fluxes might allow one to derive even tighter
constraints on parameters and eases the grounds for a self-calibration analysis. 

On the basis of the cluster flux selection function and the flux to mass
conversion, the mean expected cluster number of the fiducial models is compared with theoretical
predictions of models by using a Poisson-averaged likelihood in our MCMC
analysis (also accounting for the (small) cluster candidate sample impurity which
otherwise slightly biases the cosmological parameter
constraints\footnote{Note that in addition to predictions about the sample
  purity gained from
  simulations as described in this work, false detections will be exposed
  by follow-up observations, which need to be carried out to estimate cluster
  redshifts. Nevertheless, in order not to waste valuable observing time, a
  high purity of the chosen candidate sample is absolutely
  necessary.}). 
 Here, we want to emphasize that the selection functions (depending on
cluster flux and redshift) for the performed extraction algorithms are more complex
than it has often been assumed in previous analyses of other authors. Most
often simple step and symmetric selection functions have been applied in those
studies. In the following, the found two dimensional (redshift and flux) selection of the MFMF cluster
extraction method is used to derive cosmological parameter constraints. 
 Note further that the choice of the parameterised cluster template can affect the cluster
selection and as well bias cluster photometry and thus cosmological constraints. A discrepancy of the presumed
template from the average universal one of real clusters causes a reduction in
the cluster detection efficiency. 
Generally, in addition to ignored sources of confusion, the selection function
is mostly affected by the choice of the template and its
parameterisation. In the case of spatially highly resolved
multi-frequency data, a high adaptability of the parameterised template is advantageous.
Here, the same cluster template has been used for the SZ simulations and as
the detection template.
The detection efficiency is thus optimal.
In reality, it will be an iterative process to match the parameterised
template and its parameter priors to the profiles of real clusters.
It would be a good test of the cluster extraction algorithm to apply it to Planck data whose
SZ component has been realised by hydrodynamical N-body
simulations. However, at present such simulations of cosmic volumes and in
quantities as needed to give
robust predictions of the cluster selection for the Planck cluster survey
are not available.

\begin{figure}
\begin{center}
\epsfig{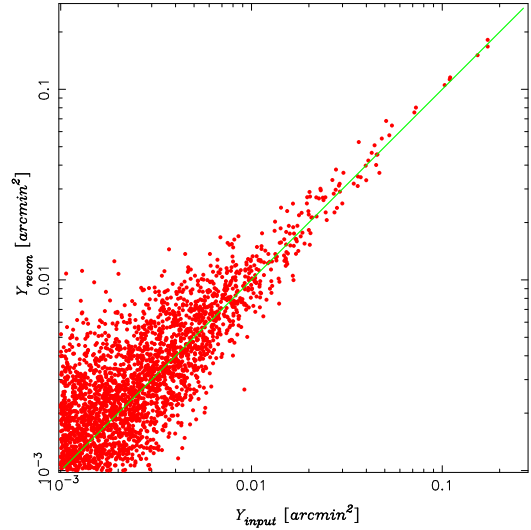}
\caption{Scatter of the reconstructed cluster fluxes ($Y_{\rm recon}$) versus the real cluster
  fluxes ($Y_{\rm input}$) of cluster candidates which are matched up with a simulation input
  cluster by the described matching algorithm. The scatter is shown for fluxes above
  $Y_{\rm cut}=1\times10^{-3}$ arcmin$^2$.}\label{fig:planckYscat}
\end{center}
\end{figure}

Furthermore, the conversion used between the real cluster
flux and the cluster mass is taken to be universal and assumed to
be free of any dispersion.
In principle there exists intrinsic scatter in the mass-flux relation due to
differences in cluster environments and evolution histories. 
 It is possible to include such an uncertainty in the relation by a
convolution of the mass function. 
 However,
since hydrodynamical cluster simulations support the assumption that a tight correlation between
the cluster mass and flux exists (see e.g. \cite{2004daSilvaetal}; \cite{2005Motletal}), we neither include a dispersion of the
scaling relation in the SZ simulations nor in the following analysis of cosmological parameter
constraints.
Estimates of the scatter intrinsic to the mass-flux relation obtained from
numerical cluster simulations determine it to be of the order of a few percent.
In the case one wants to make use of the SZ cluster flux function ($\log
S - \log N$) by binning
clusters according to their reconstructed fluxes, it is in general
not the intrinsic scatter in the mass-flux relation which causes the
largest uncertainty of the number of clusters contained in a flux bin.
It is rather the scatter of the reconstructed fluxes to the real cluster
fluxes that prevents one from finely flux binning 
 the recovered
sample. As it can be seen from the flux scatter plots shown
in Figure \ref{fig:planckYscat} and Figures 9, 10 and 12 of GKH05, which are comparable to the
one of the algorithm employed in this work,
the uncertainty in the relation of the reconstructed to the real cluster flux
exceeds by far for most flux limits except for the highest flux clusters on the
sky the intrinsic scatter of the mass-flux relation of individual clusters. 

\subsection{Results}

\begin{figure*}
\begin{center}
\epsfig{file=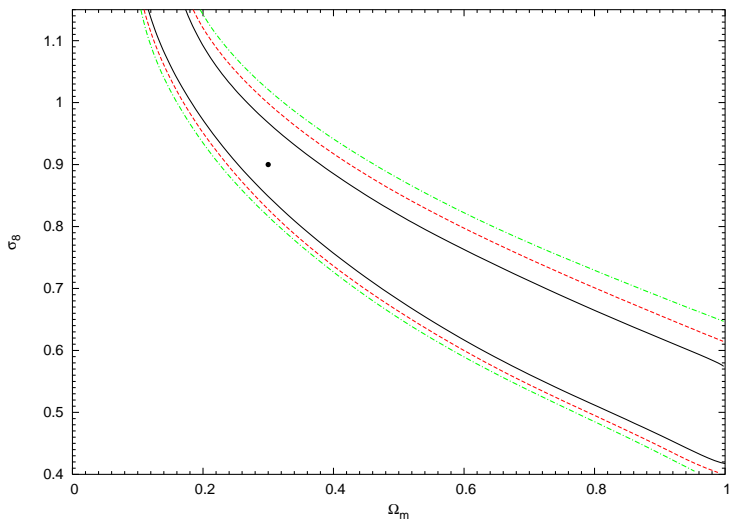,width=7.cm,angle=-90,clip=}
\caption{Cosmological parameter constraints on $\Omega_m$ and $\sigma_8$ from
  total cluster number counts on the full sky and without cluster redshift
  information at hand.
  The 68.3\% (black solid line), 95.4\% (red dashed line) and 99.7\% (green
  dash-dotted line) confidence levels are shown. The two parameters are
  degenerate for total number counts. Without prior knowledge by other means only the shape
  of the degeneracy relation can be constrained. A liberal top-hat prior is
  placed on the Hubble constant ($0.4 \leq h \leq 1$). As fiducial cosmological model
  the concordance model has been assumed ($\Omega_m = 0.3$, $\Omega_\Lambda =
  0.7$, $\sigma_8 = 0.9$ and $h = 0.7$).}\label{fig:planckclsurvconstrnored}
\end{center}
\end{figure*}

In the following results on parameter constraints are given for several
assumptions concerning the restrictions placed on parameters by priors
(notably restrictions on the Hubble parameter, $h$) and for different degrees of
effort of following-up the cluster sample in the optical for providing cluster
redshifts (actually we distinguish between no and complete follow-up). 
In doing so, we start from weak assumptions and assume an increase in
knowledge about the sample redshifts as well as 
 tighten the prior on $h$. These actions lead, as expected, to
ever tighter constraints on cosmological parameters obtainable from the
cluster sample.

Due to the redshift independence of the SZ effect and due to the resolution of
the Planck channels, cluster redshift information cannot be gained from Planck
data on its own. In the case of high resolution SZ observations, redshifts
can be estimated based on morphological observables (see \cite{2003Diegoetal}).
However, the error bars given by the morphological redshift estimation are
large. Still such redshift determination might be useful for future SZ
surveys, such as the SPT, which will presumably detect thousands of clusters.
Obtaining redshifts for such a large number of clusters by follow-up
observations at different wavelengths
represents currently a major
observational effort. Therefore, we first examine the case that no cluster redshift
information is available. Figure \ref{fig:planckclsurvconstrnored} shows the `constraints' that can be
obtained from the total number of SZ detected clusters for $\Omega_m$ and
$\sigma_8$. As expected the two parameters are completely degenerate. One can
always find a combination of these two parameters, which mainly govern the mass
function, that reproduces the
observed total cluster number on the full sky. 
The particular shape of the degeneracy depends on the survey layout, e.g. the
survey depth that can be reached. The degeneracy between $\Omega_m$ and
$\sigma_8$ can be broken somewhat by utilising angular correlation function information (see
\cite{2004MeiBartlett}). The `width' of the degenerate constraints at a given
(fixed) value of $\Omega_m$ or $\sigma_8$
respectively depends strongly on the range of the $h$ prior. Here we used a
uniform prior on $h$ with $0.4\leq h\leq 1$. Another way to break the degeneracy
between $\Omega_m$ and $\sigma_8$ and to constrain $h$ at the same time is to
carry out a combined data analysis utilising the Planck cluster sample and the
primordial CMB power spectrum (or temperature and polarisation power spectra)
obtainable from Planck data.

In our first analysis only the $\Lambda$CDM concordance model ($\Omega_m = 0.3$, $\Omega_\Lambda =
  0.7$, $\sigma_8 = 0.9$ and $h = 0.7$) has been
considered as fiducial model. Since the recently published results of the
(primordial) cosmic microwave background analysis of the WMAP three year
data prefer different parameters than the concordance model, we include apart
from the concordance model the best fit WMAP cosmology ($\Omega_m = 0.27$, $\Omega_\Lambda =
  0.73$, $\sigma_8 = 0.75$ and $h = 0.7$) in our examinations and
compare the constraints on the parameters of the two models. In particular, the
change in $\sigma_8$ has an influence on the expected number of clusters
recoverable from Planck data.
  
In order to estimate by how much constraints on cosmological parameters are improved
by provided cluster redshift information, we further (optimistically) assume
that cluster redshifts are known within some error for the complete Planck
cluster sample.
This raises questions about the feasability of obtaining cluster redshifts for
a major fraction or for the entire sample respectively.
The presently least expensive way to measure cluster redshifts is by
performing multi-band near-IR and optical imaging observations.
First, let us discuss how likely it will be at the time when Planck will have
collected its data to have access to optical data for redshift estimations as needed here.
With surveys, such as the Sloan Digital Sky Survey (SDSS)\footnote{\texttt{http://www.sdss.org}} and the Two-degree
Field Galaxy Redshift Survey (2dFGRS)\footnote{\texttt{http://www.mso.anu.edu.au/2dFGRS}} being already in place and even larger
surveys being funded and becoming operational at about the time Planck
finishes data collection (e.g. Pan-STARRS, expected to start scanning the sky in 2010), it is
not too far-fetched to assume that for a major fraction of the sky optical
data of high quality will be available.
This assumption of having redshift information available for a large number of
Planck clusters is further supported by the fact that the median redshift
of the Planck cluster sample of $z \approx 0.2$ matches reasonably well the median
redshifts of today's large scale galaxy surveys (i.e. the galaxy samples of the
SDSS have median redshifts of $z \approx 0.104$ (main galaxy sample) and $z
\approx 0.35$ (luminous red galaxies) and that of the 2dFGRS is
$z \approx 0.11$).

\begin{figure*}
\begin{center}
\epsfig{file=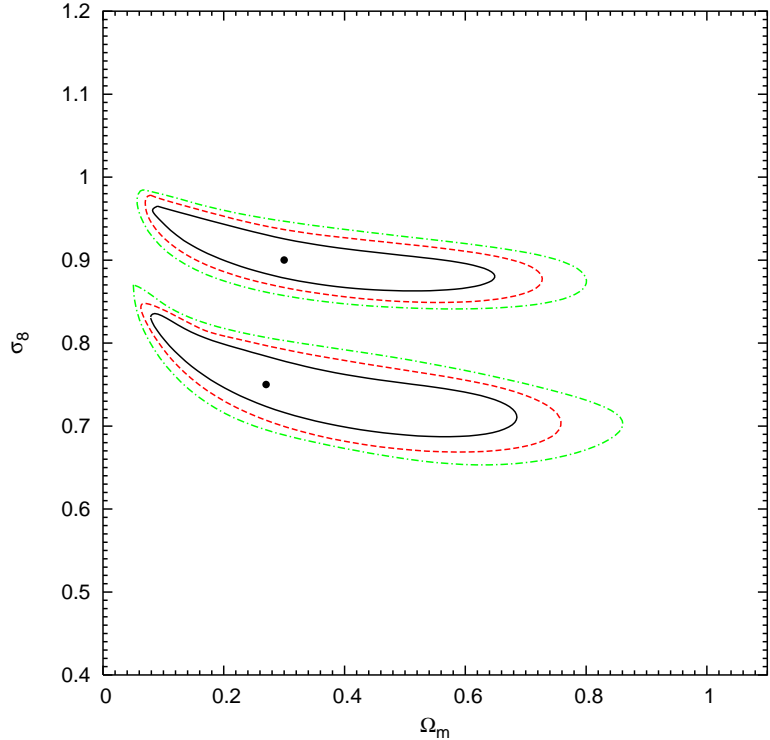,width=10.cm,angle=-90,clip=}
\caption{Cosmological parameter constraints on $\Omega_m$ and $\sigma_8$ from
  the full sky Planck survey cluster sample. }\label{fig:2dOmsig8concordwmap}
\end{center}
\end{figure*}

Furthermore, as the Planck sample contains only (very) massive clusters (commonly
$M_{\rm cl} \gtrsim 5\times 10^{14}$ $h^{-1}M_{\odot}$), clusters of the
sample will have a high richness (number of member galaxies) and contain many
bright galaxies. Further, since the massive clusters of the sample should
exhibit a strongly developed red-sequence of galaxies, a deep
two-band photometry might be an economical way to gain cluster redshifts of
the precision needed for clusters located on the sky remotely from
covered optical survey areas.
Nevertheless, due to the large uncertainties of the cluster position caused by the rather
coarse Planck channel resolutions, pointed follow-up observations of single
clusters might be a difficult and cumbersome undertaking.
This complicates pointed follow-up observations in optical as well as X-ray
wavebands (for a detailed discussion on Planck cluster sample follow-up at
different wavelengths and on the expected properties of clusters detectable by
Planck in other wavebands see \cite{2003White}).
Even in the case of large surveys at optical and/or X-ray wavebands, the recovered
positions of Planck detected clusters give only weak constraints for
locating associated cluster characteristic features in data
collected over wide fields at these wavebands.
A similar procedure of matching up Planck detected clusters with detections
at other wavebands, as described in section \ref{sec:tszmfmfplanckcat}, has
rather to be adopted, after cluster
candidates have been located within a matching region at the respective other
waveband under consideration.
For the case of finding X-ray counterparts, the RASS
(\cite{1991Truemper}; \cite{1999Vogesetal}) is a good base for providing
matches for Planck clusters with redshifts of $z\lesssim 0.3$. The main
existing X-ray instruments used these days to detect clusters, XMM-Newton and
Chandra, may be not operational anymore at the time when Planck completes its
data collection. However, there should be a
large overlap between a combined catalogue of cluster detections made by them
and the future Planck sample. 

Since the Planck cluster sample consists mainly of low redshift massive
clusters (the Planck sample is unlikely to contain detections with redshifts
$z \gtrsim 1$ for the fiducial cosmological models), 
 already
quite shallow multi-band optical surveys covering a wide field (as the ones
mentioned above) are well sufficient for cluster redshift determination.
Having data of up-coming surveys, such as Pan-STARRS and the Large Synoptic
Telescope, available eases the redshift hunt even further.
In the following, we use a conservative redshift binning of $\Delta
z=0.1$ to group clusters in redshift bins. Ideally one would like to have access to
spectroscopic redshifts whose $1\sigma$ precision is commonly better than $\Delta
z=0.01$ for an individual galaxy in the redshift range of interest. However,
availability of spectroscopic redshifts is likely to be limited. Nevertheless,
photomotric redshifts suffice
as well for our purposes. For example, photometric redshifts derived from the
five SDSS bands are accurate to $\Delta z \sim 0.03$ for an individual
galaxy. For a
massive cluster hosting several detectable galaxies ($N_{\rm gal}$) the photometrically
determined redshift estimate precision increases proportional to $\sqrt{N_{\rm
 gal}}$. Even for a reasonably deep two-band photometry which uses the 4000 angstrom
break of the red cluster sequence galaxies, cluster redshift estimates
obtained from colour-magnitude diagrams have uncertainties below $\Delta z \sim
0.05$ for most of the considered redshift range.
Therefore, with cluster redshift determination on photometric grounds,
redshift bins of $\Delta z=0.1$ are well feasable.
Furthermore, our presumed choice of redshift binning is optimal in this
respect that it avoids significant cross-correlations between seperate
adjacent bins. Covariances between them can therefore be neglected in the
analysis. 

At first, we derive constraints assuming a very weak prior on the Hubble
parameter: $0.02 \leq h \leq 5$ (actually this corresponds to $h$ being
unconstrained). The resulting obtainable constraints on $\Omega_m$ and $\sigma_8$ are shown in
Figure \ref{fig:2dOmsig8concordwmap} for the two fiducial cosmologies without requiring
the geometry of the Universe to be flat. Since
the detected cluster number is statistically relevant, due to the
lower number of detected clusters in models with low values of $\sigma_8$, as
expected, the
constraints derived for the WMAP fiducial model are weaker than the ones on
the concordance model parameters. While the analysis is able to
place reasonable constraints on $\sigma_8$ for both fiducial models ($\Delta
\sigma_8 \lesssim 0.08$ at all times), it is
barely feasable to gain useful restraining information about the matter
density $\Omega_m$. However, an Einstein-de Sitter model ($\Omega_m=1$) can be
excluded at high significance in both cases. Note that on the basis of the
performed analysis the two fiducial models exclude each other at several ($\gg
3$) $\sigma$.

\begin{figure*}
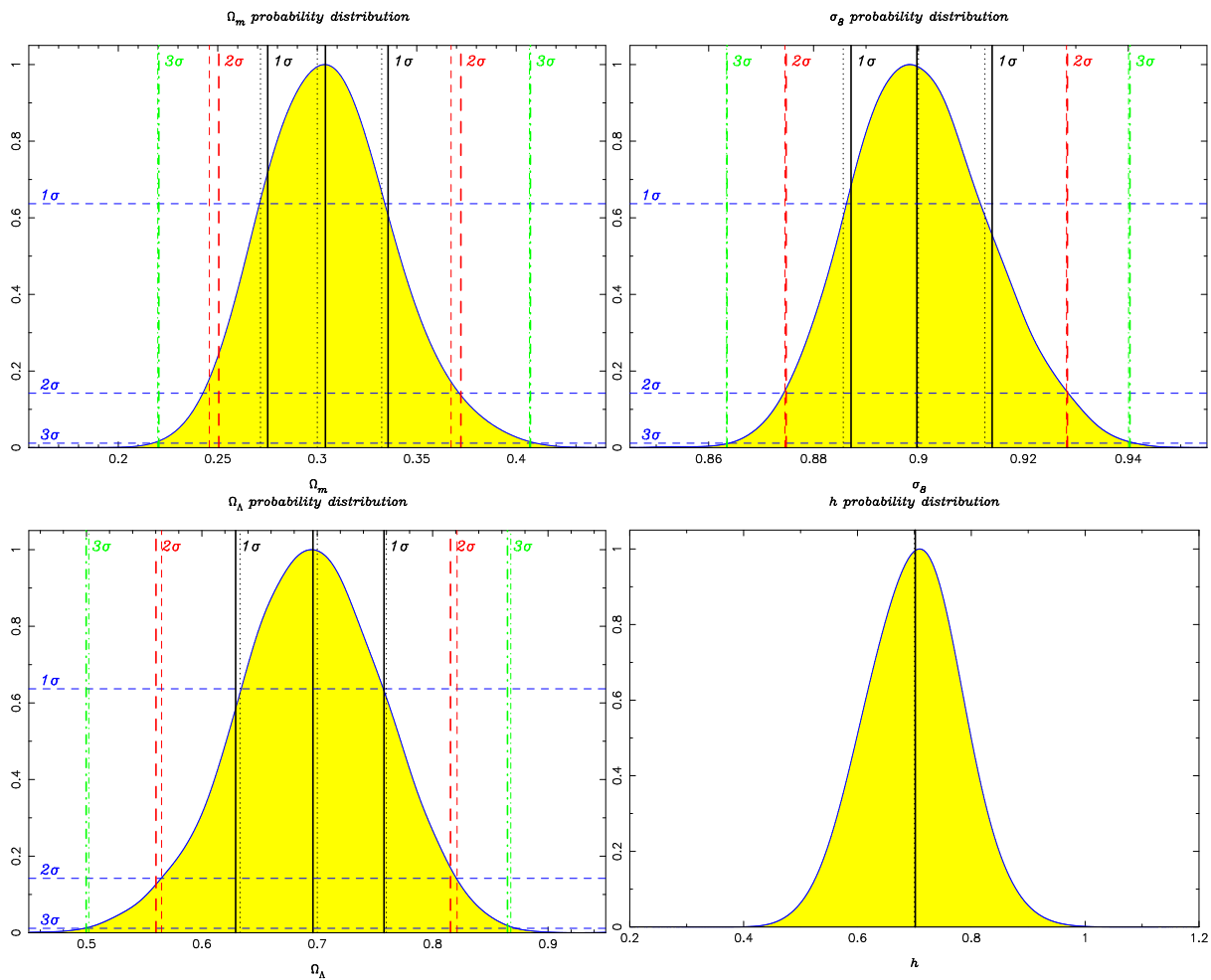

\begin{center}
\includegraphics[angle=-90,width=0.45\textwidth]{figs/planck_cospar/planck_Om.ps}
\includegraphics[angle=-90,width=0.45\textwidth]{figs/planck_cospar/planck_sig8.ps}
\includegraphics[angle=-90,width=0.45\textwidth]{figs/planck_cospar/planck_OL.ps}
\includegraphics[angle=-90,width=0.45\textwidth]{figs/planck_cospar/planck_h.ps}
\caption{One-dimensional marginalised probability distributions for the
  cosmological parameters varied in our MCMC analysis. 
  There has been no direct restriction on the
  curvature $\Omega_k=1-\Omega_m-\Omega_\Lambda$ (no flatness
  prior). 
  Apart from the Hubble parameter, the remaining parameter sub-space of the
  other parameters is sampled uniformly. Note that redshift
  number counts of cluster surveys are most sensitive to $\sigma_8$ and in the
  case of a tight prior on $h$ constrain $\Omega_m$. The marginalised
  distributions are obtained from thinned MCMC sample chains. Thinning ensures
  that chain samples used for estimating parameter confidence intervals are
  de-correlated. Chains are thinned in such way that the correlation between
  consecutive chain steps in the effective thinned chains are below $0.5$
  when defining the correlation to be unity at zero step size. 
  All over several million samples
  are taken to explore the likelihood distribution. We refer the reader to the text
  for an explanation of the different confidence intervals shown in the
  panels. Note that, even though a large number of samples has been taken, the
  error on the 99.7\% confidence level of each parameter is rather large due
  to a still small number of samples located outside this confidence level
  and thus a resulting large relative sample variance. \label{fig:onedimplanckcosconstr}}
\end{center}
\end{figure*}

\begin{figure*}
\begin{center}
\includegraphics[angle=-90,width=0.45\textwidth]{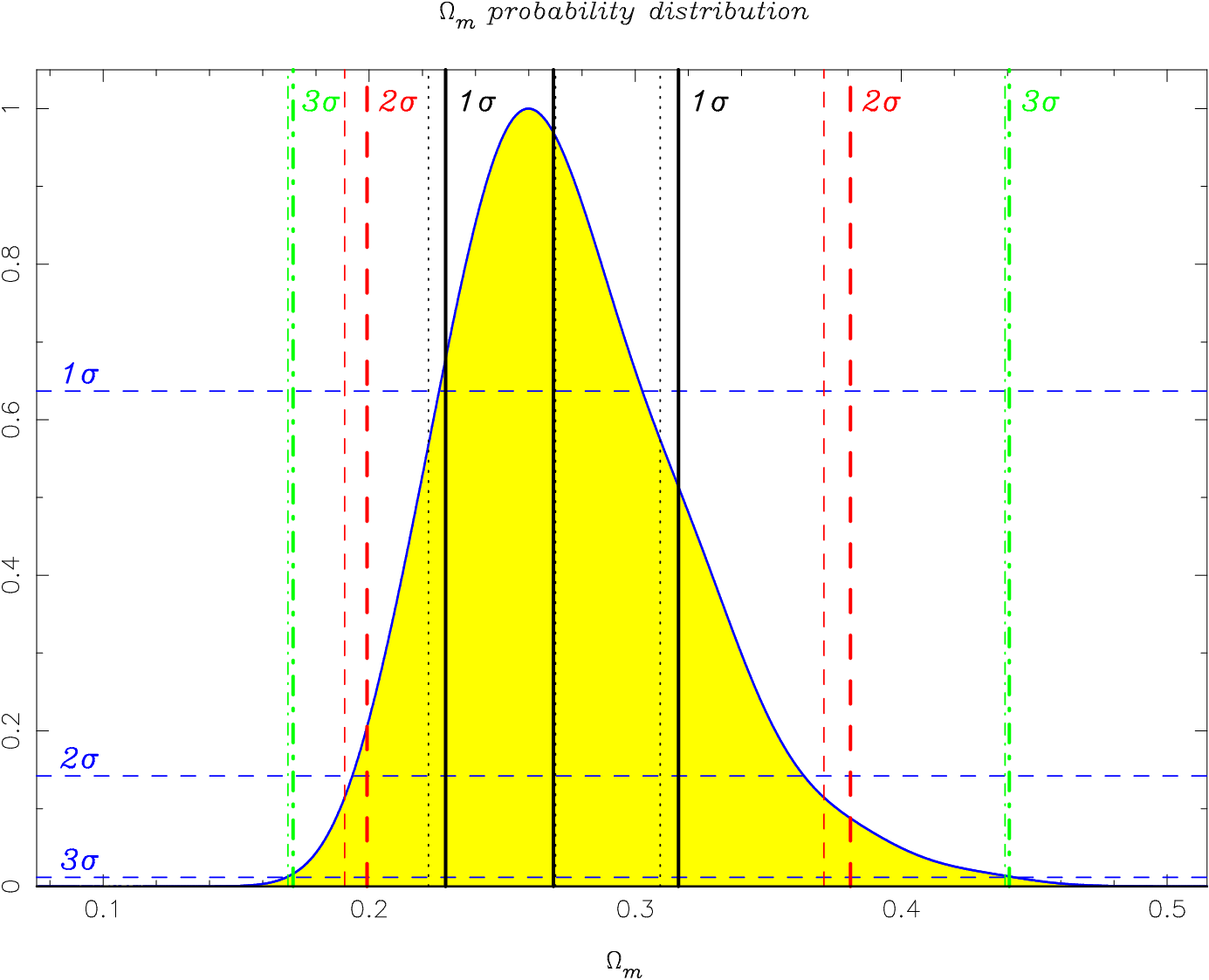}
\includegraphics[angle=-90,width=0.45\textwidth]{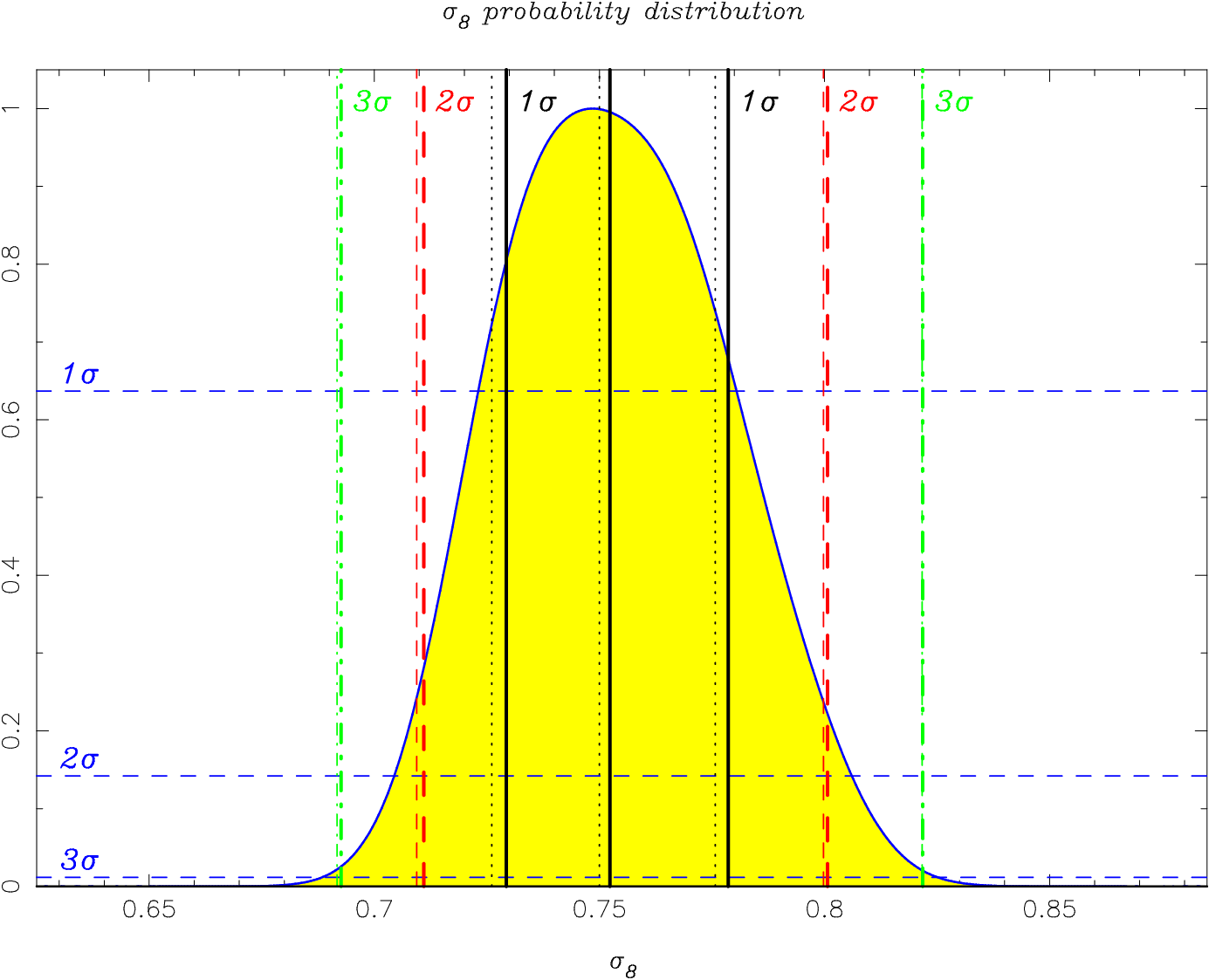}
\includegraphics[angle=-90,width=0.45\textwidth]{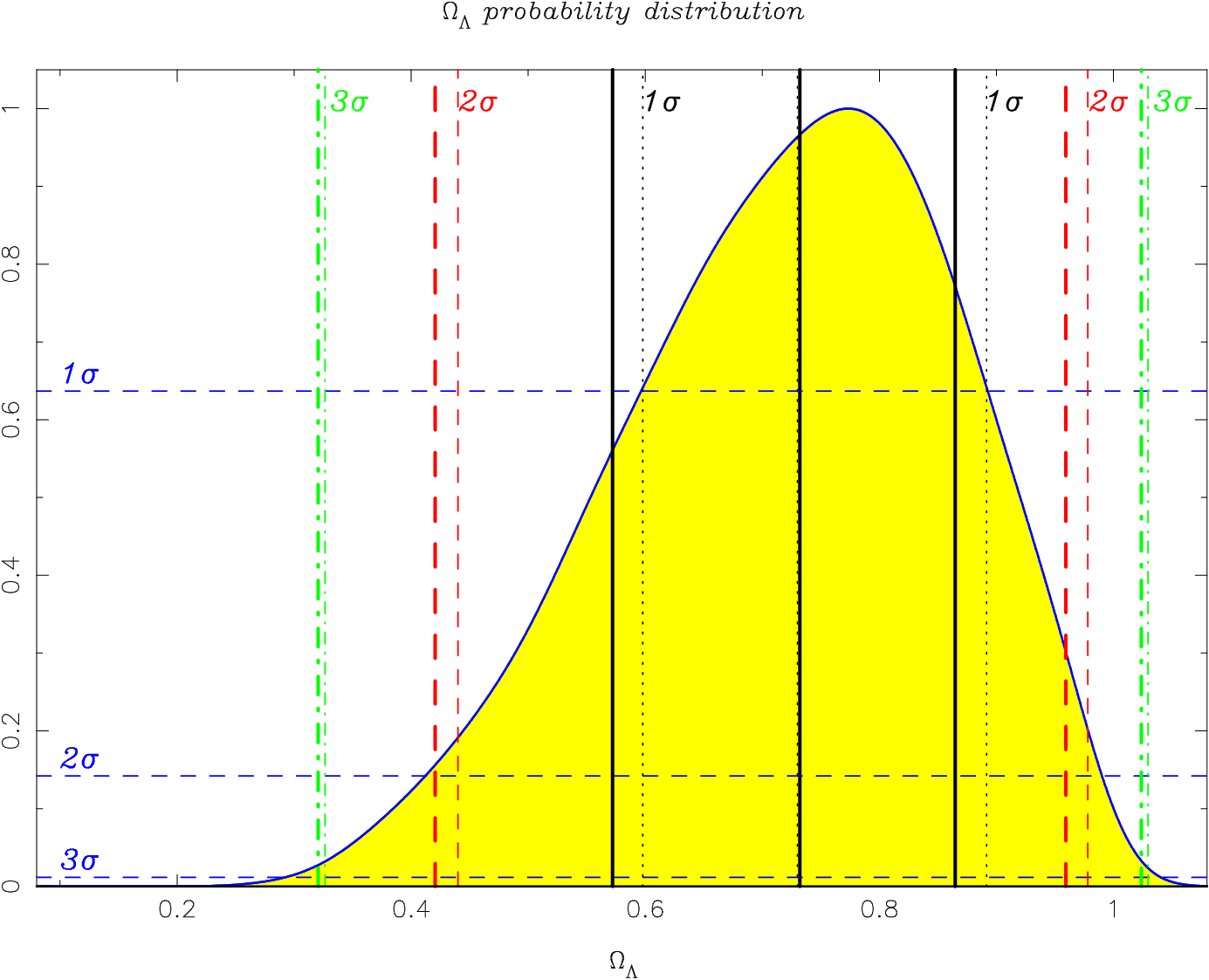}
\includegraphics[angle=-90,width=0.45\textwidth]{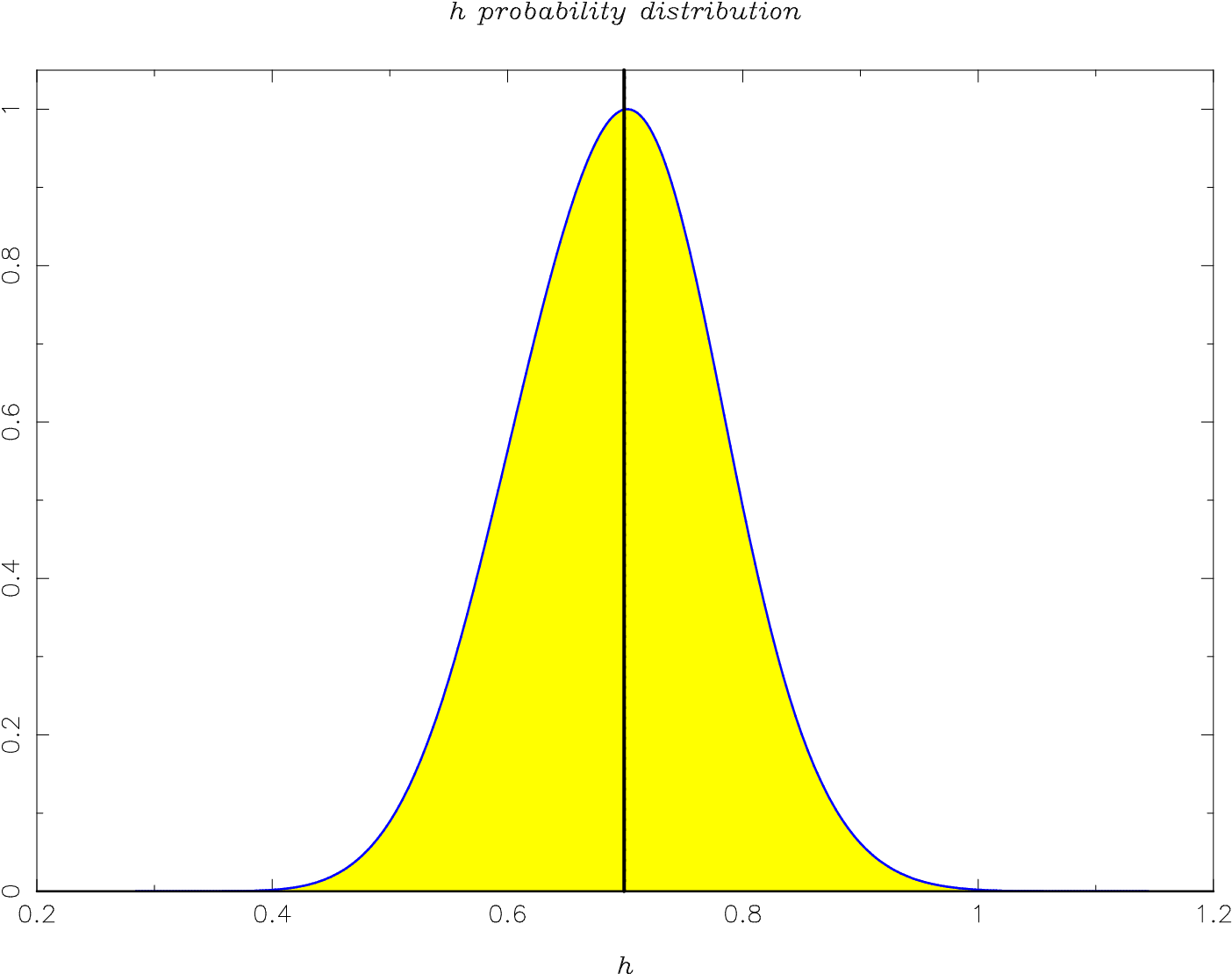}
\caption{The same as Figure \ref{fig:onedimplanckcosconstr} for the WMAP best
  fit fiducial cosmological parameter model ($\Omega_m=0.27$,
  $\Omega_\Lambda=0.73$, $\sigma_8=0.75$ and $h=0.7$).\label{fig:onedimplanckcosconstrwmap}}
\end{center}
\end{figure*}

Finally, we place a tight prior on $h$. For the further analysis
we constrain the Hubble parameter to $h=0.7\pm0.08$, as supported by the HST
Key Project (\cite{2001Freedmanetal}). 
Figure \ref{fig:onedimplanckcosconstr} and \ref{fig:onedimplanckcosconstrwmap}
show the marginalised one-dimensional likelihood distributions of the four
cosmological parameters, which have been allowed to vary in our analysis, for
the two fiducial cosmologies. In addition to the central expectation and fiducial
parameter values, various confidence intervals are plotted as well.
The dotted central vertical line in each panel indicates the fiducial
parameter value. The central thick solid line gives in each case the estimated
parameter value gained from the MCMC analysis. Hereby, the shown parameter estimate
corresponds to the median of the particular distribution.
 The other vertical lines give the
quantiles of the distributions that are used to quote confidence limits on the
parameter constraints. In each case confidence interval lines indicated by
different colours and linestyles enclose 68.3\% (black solid/dotted), 95.4\%
(red dashed) and 99.7\% (green dot-dashed)
confidence regions. Next to the respective confidence interval lines the
corresponding confidence level is given in the same colour. The thin lines
correspond to quantiles 
 which enclose a particular percentage of the samples of the contributing
 chains by intersecting the likelihood distribution at the same
`height' on both sides of the distribution peak.
Thus the area under the graph outside the interval limits adds, for example,
(asymmetrically for a non-Gaussian distribution) up in total to 31.7\% of the
entire area under the graph in the case of the 68.3\% confidence
region. However, both sides do not have to contribute the same area.
The thick lines represent confidence limits obtained from restricting in each
case the
integrated areas under the graph to be the same on both sides of the
distribution peak.
This means that for the $1\sigma$ confidence limit interval, 15.85\% of the
samples of the chains have values below the lower confidence limit and the
same percentage of sample values lie above the upper confidence limit.
Note that there is no assumption about the distributions being Gaussian.
The horizontal dashed blue lines indicate $\exp{(-x^2/2)}$ for $x=1$, 2 and 3
respectively. In the case a probability distribution is Gaussian, the
intersections of the distribution with the respective line correspond to the $1\sigma$,
$2\sigma$ and $3\sigma$ confidence limits respectively. Therefore, if the
vertical and horizontal lines do not cross each other at the point at which they
intersect the distribution graph, the distribution is non-Gaussian. Moreover, for a
Gaussian likelihood distribution the two different ways of placing
confidence limits (thin and thick lines) agree.
Besides for the Hubble parameter, the marginalised distributions for the most part deviate
from a Gaussian one. Even the likelihood distribution of the Hubble
parameter which is a priori restricted to a Gaussian is marginally
screwed which is caused by degeneracies between $h$ and other variable
parameters in the analysis.

The parameter expectation values (median of the particular distribution)
gained from the MCMC likelihood analysis
together with the 68.3\% confidence limits are listed in Tables
\ref{tab:planckcosconstr} and \ref{tab:planckcosconstrwmap} for the fiducial models.
By comparing the parameter constraints of Tables \ref{tab:planckcosconstr} and
\ref{tab:planckcosconstrwmap}, it can be seen, parameters are tighter
constrained (confidence intervals are smaller) in the case of the concordance
model. This is due to the higher number of detectable clusters in the
concordance model. 

Tight constraints are especially placed on the variance of matter fluctuations on scales of $8
h^{-1}$Mpc, $\sigma_8$. This is even the case for a less restrictive $h$ prior. 
Under the made assumptions and set priors (fixed $n_s$ and $\Omega_b$ and
prior on $h$), Planck cluster redshift number counts surpass recent primordial CMB
power spectrum measurements in the ability to constrain
$\sigma_8$. However, loosening the restrictions on the spectral index $n_s$ and
the baryon density $\Omega_b$ weakens constraints on $\sigma_8$. 
On the other hand primordial CMB power spectrum evaluations are well
suited to constrain $n_s$, $\Omega_b$ and $h$. Thus, a combined analysis of
the Planck cluster sample and the primordial CMB power spectrum recovered from
Planck CMB data is downright recommended. Moreover, combining cluster number
counts with investigations of angular (possibly spatial)
clustering 
of the galaxy clusters in the sample and estimates of their gas (baryonic) mass
fraction from multi-waveband observations may as well result in a further
improvement of constraints which are 
 based 
on the Planck cluster sample and its follow-up observations.

Furthermore, it is also feasable to derive tight constraints on the matter density
parameter $\Omega_m$ if a restrictive $h$ prior is set (see Tables \ref{tab:planckcosconstr} and \ref{tab:planckcosconstrwmap}). 
The constraining power of the Planck cluster sample on $\Omega_m$ is
comparable to the one obtained from the three year WMAP data alone in an
analysis with six free parameters assuming the Universe to be flat (see \cite{2006Spergeletal}).
Even the dark energy content $\Omega_\Lambda$ and the curvature $\Omega_k$ can
be constrained. However, they are the ones which of all the parameters in our analysis
are least constrained. The 68.3\% confidence limits for $\Omega_\Lambda$ are given
in Tables \ref{tab:planckcosconstr} and \ref{tab:planckcosconstrwmap}. By
adding $\Omega_m$ and $\Omega_\Lambda$ of each MCMC
chain sample the scatter of the sample curvature values around flatness can be
investigated for each fiducial (flat) model
since we have not placed priors in our analysis on the geometry of the
Universe. Hence, one obtains: $\Omega_k=1-\left(
0.999^{+0.041}_{-0.041}\right)$ and $\Omega_k=1-\left( 0.999^{+0.101}_{-0.110}\right)$ respectively.

\begin{table}
\begin{center}
\begin{tabular}{||c|c||} \hline
Parameter & Median ($1\sigma$ constraint)\\
\hline
&\\
$\Omega_m$ & $0.304^{+0.032}_{-0.030}$\\
&\\
\hline
&\\
$\sigma_8$ & $0.9^{+0.014}_{-0.013}$\\
&\\
\hline
&\\
$\Omega_\Lambda$ & $0.697^{+0.061}_{-0.065}$\\
&\\
\hline
&\\
$h$ & $0.7^{+0.08}_{-0.08}$ (prior)\\
&\\
\hline
\end{tabular}
\caption{Derived parameter estimates from the MCMC analysis (sample distribution medians) and
  68.3\% confidence regions with interval limits given by the 15.85\% and
  84.15\% percentiles. The shown constraints are for the concordance
  $\Lambda$CDM model. \label{tab:planckcosconstr}}
\end{center}
\end{table}

\begin{figure*}
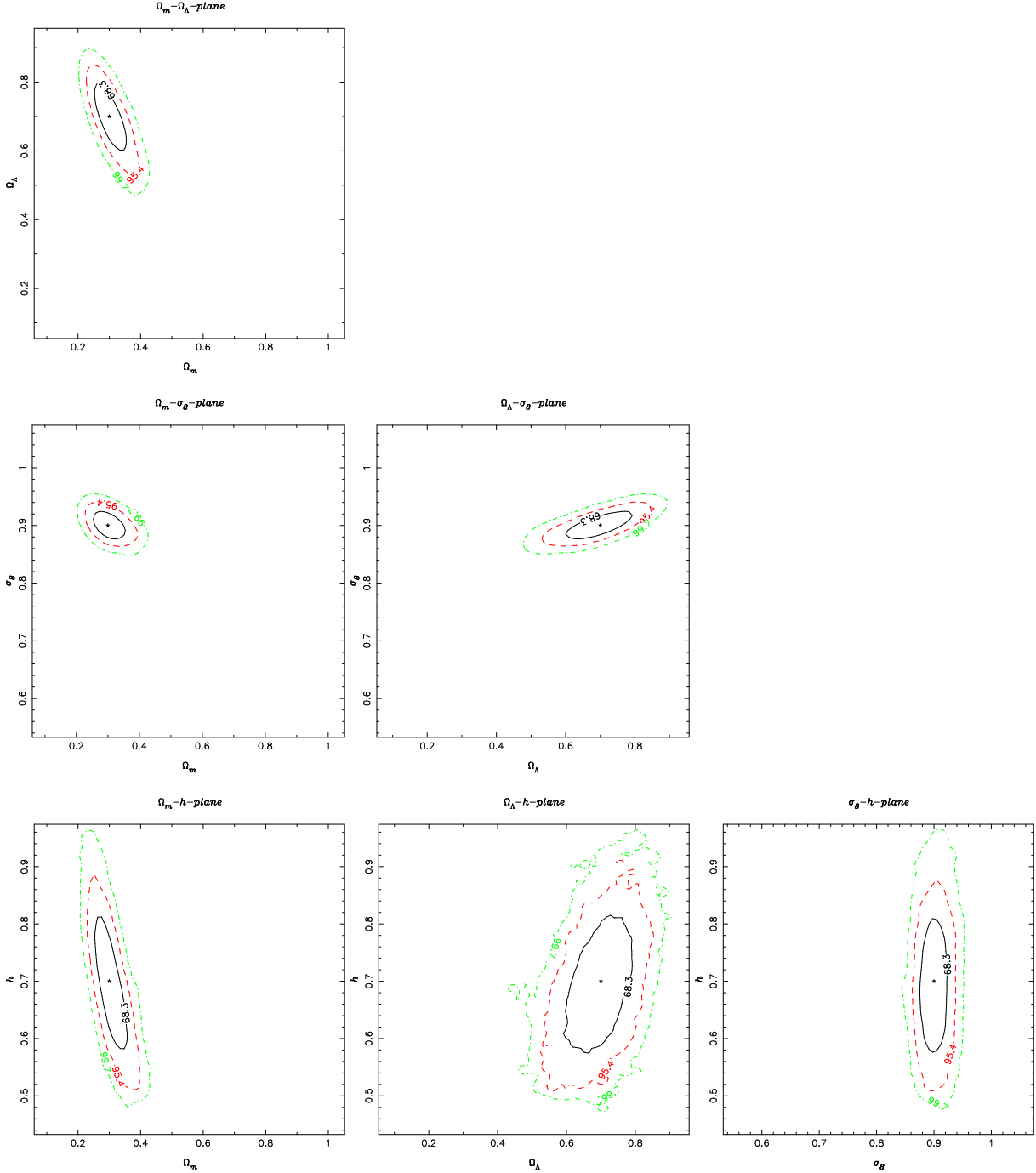

\begin{flushleft}
\subfigure{\includegraphics[angle=-90,width=0.3\textwidth]{figs/planck_cospar/concord_hprior/Om_OL_concord_hprior.ps}}\\
\subfigure{\includegraphics[angle=-90,width=0.3\textwidth]{figs/planck_cospar/concord_hprior/Om_sig8_concord_hprior.ps}}
\subfigure{\includegraphics[angle=-90,width=0.3\textwidth]{figs/planck_cospar/concord_hprior/OL_sig8_concord_hprior.ps}}\\
\subfigure{\includegraphics[angle=-90,width=0.3\textwidth]{figs/planck_cospar/concord_hprior/Om_h_concord_hprior.ps}}
\subfigure{\includegraphics[angle=-90,width=0.3\textwidth]{figs/planck_cospar/concord_hprior/OL_h_concord_hprior.ps}}
\subfigure{\includegraphics[angle=-90,width=0.3\textwidth]{figs/planck_cospar/concord_hprior/sig8_h_concord_hprior.ps}}
\caption{Two-dimensional confidence contours for all pairs of parameters. To obtain the confidence
  regions we marginalise in each case over the respective other parameters. In
  each case the contours enclose the 68.3\% (black solid line), 95.4\% (red
  dashed line) and the 99.7\% (green dot-dashed line)
  confidence regions. The underlying fiducial model is the concordence
  $\Lambda$CDM model.\label{fig:2dcont}}
\end{flushleft}
\end{figure*}

\begin{figure*}
\begin{flushleft}
\subfigure{\includegraphics[angle=-90,width=0.3\textwidth]{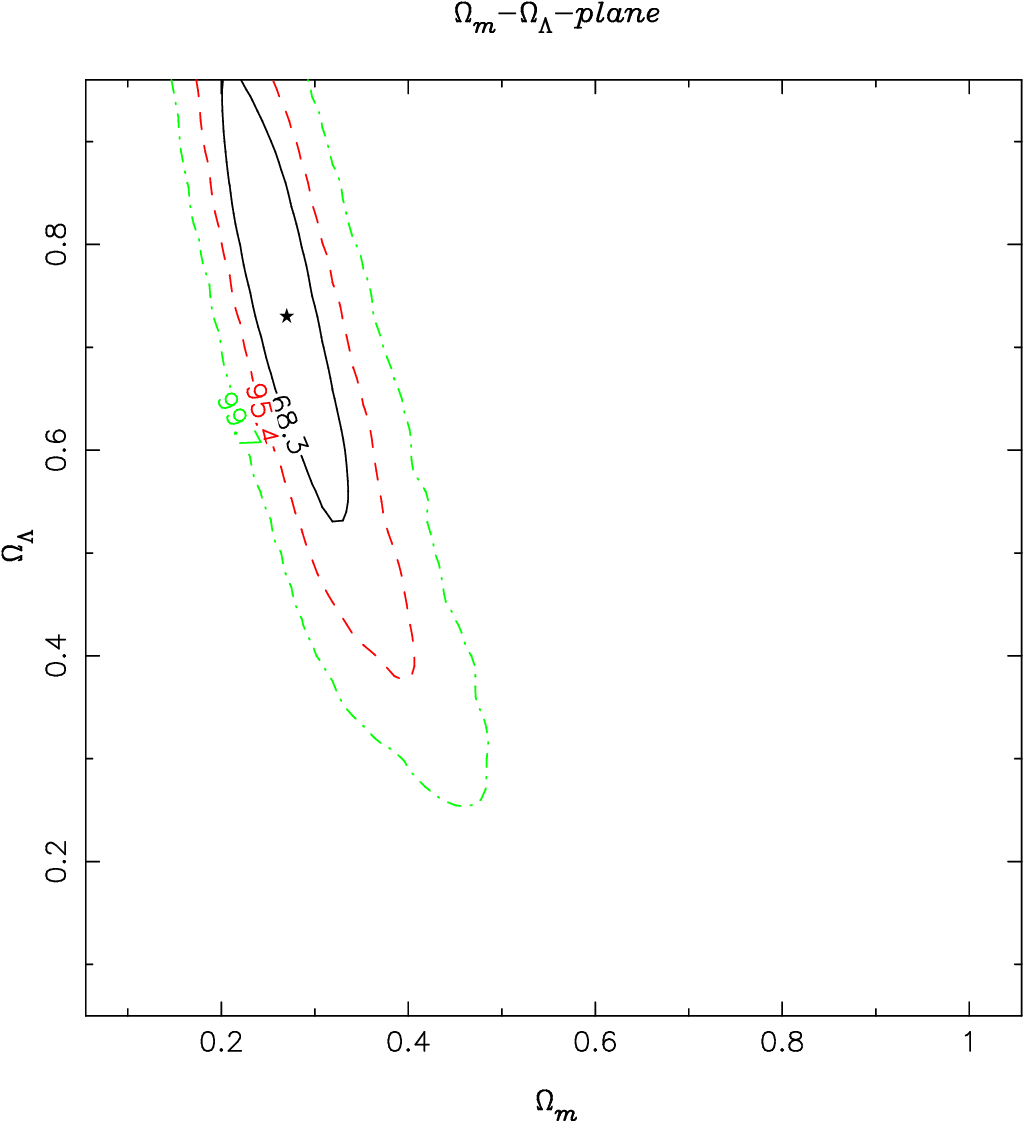}}\\
\subfigure{\includegraphics[angle=-90,width=0.3\textwidth]{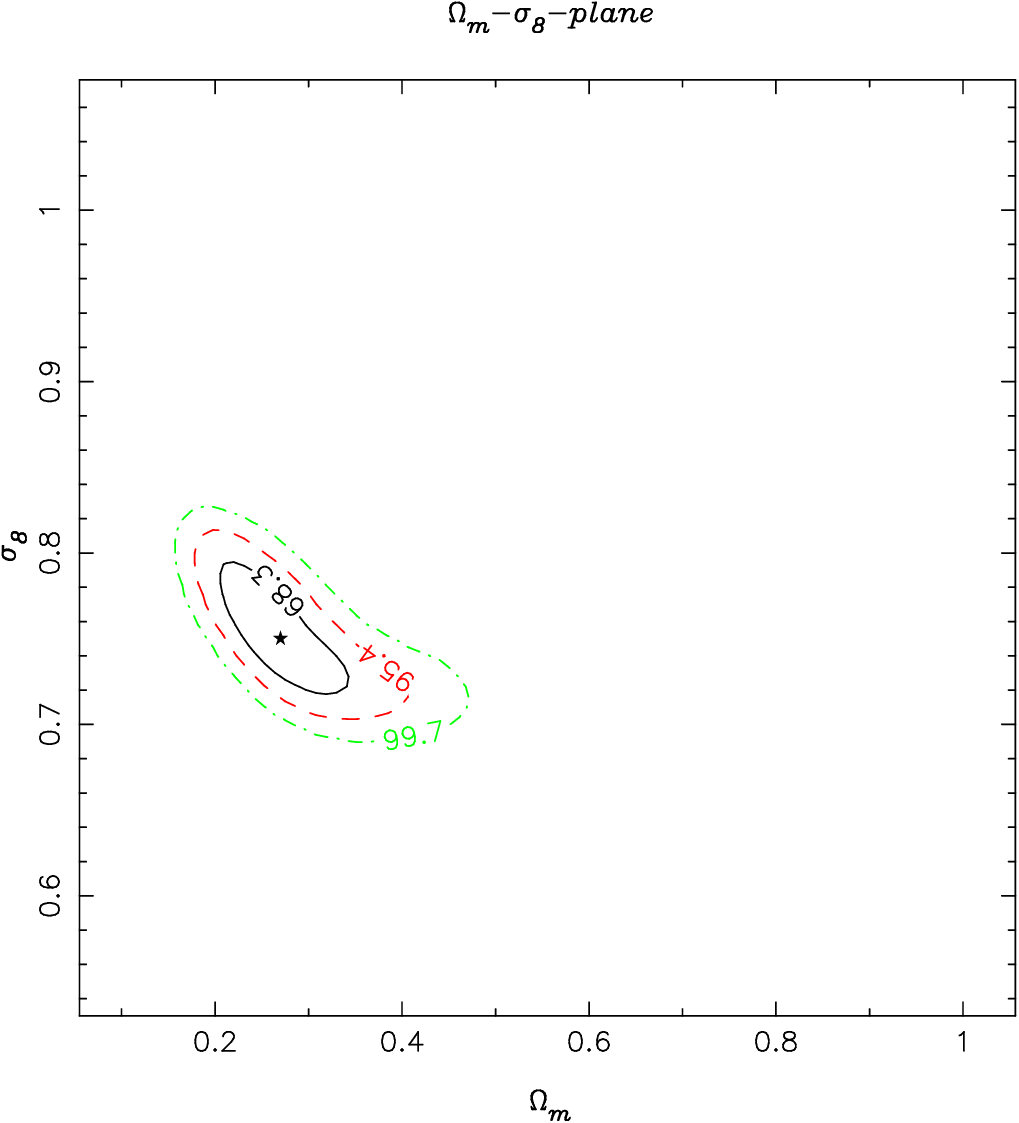}}
\subfigure{\includegraphics[angle=-90,width=0.3\textwidth]{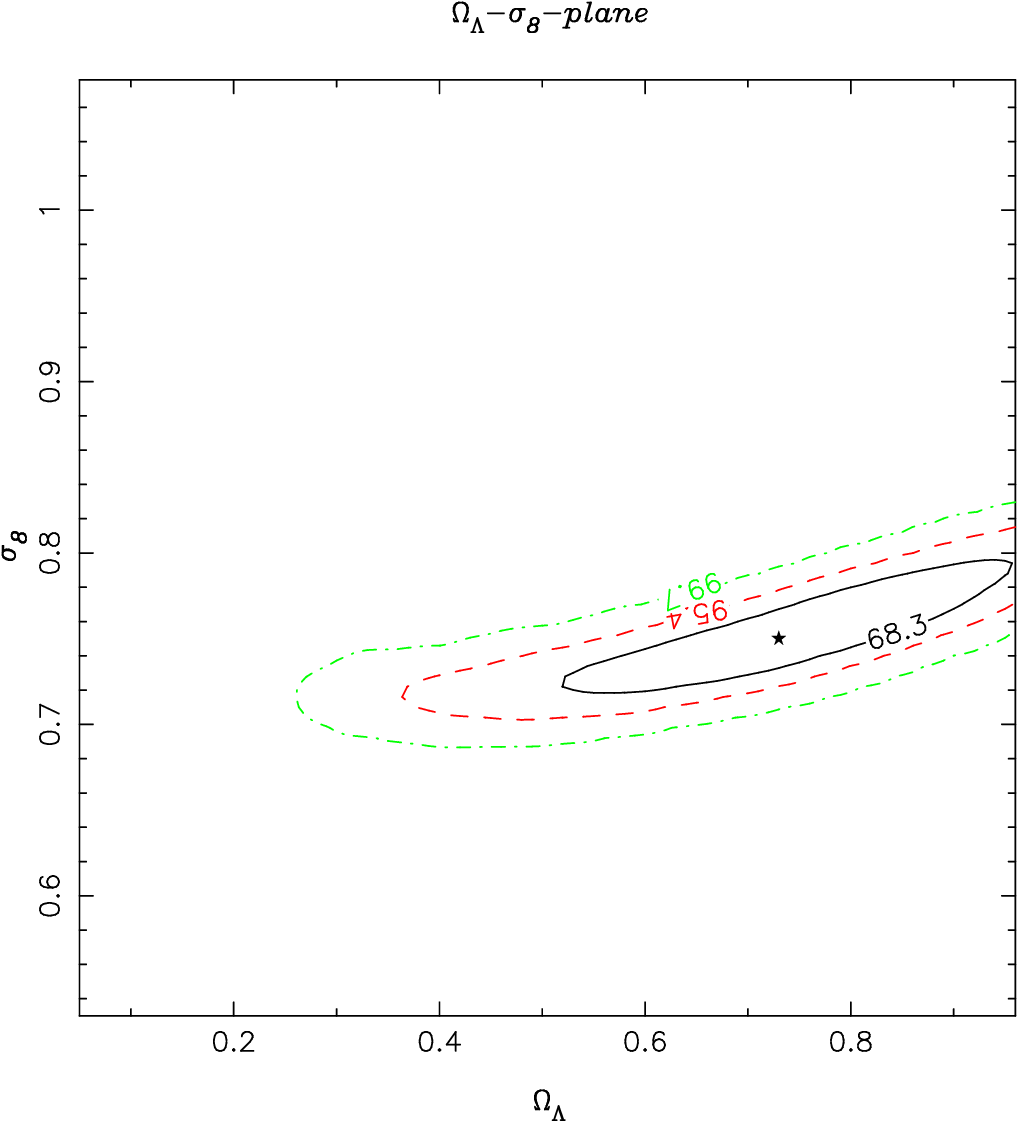}}\\
\subfigure{\includegraphics[angle=-90,width=0.3\textwidth]{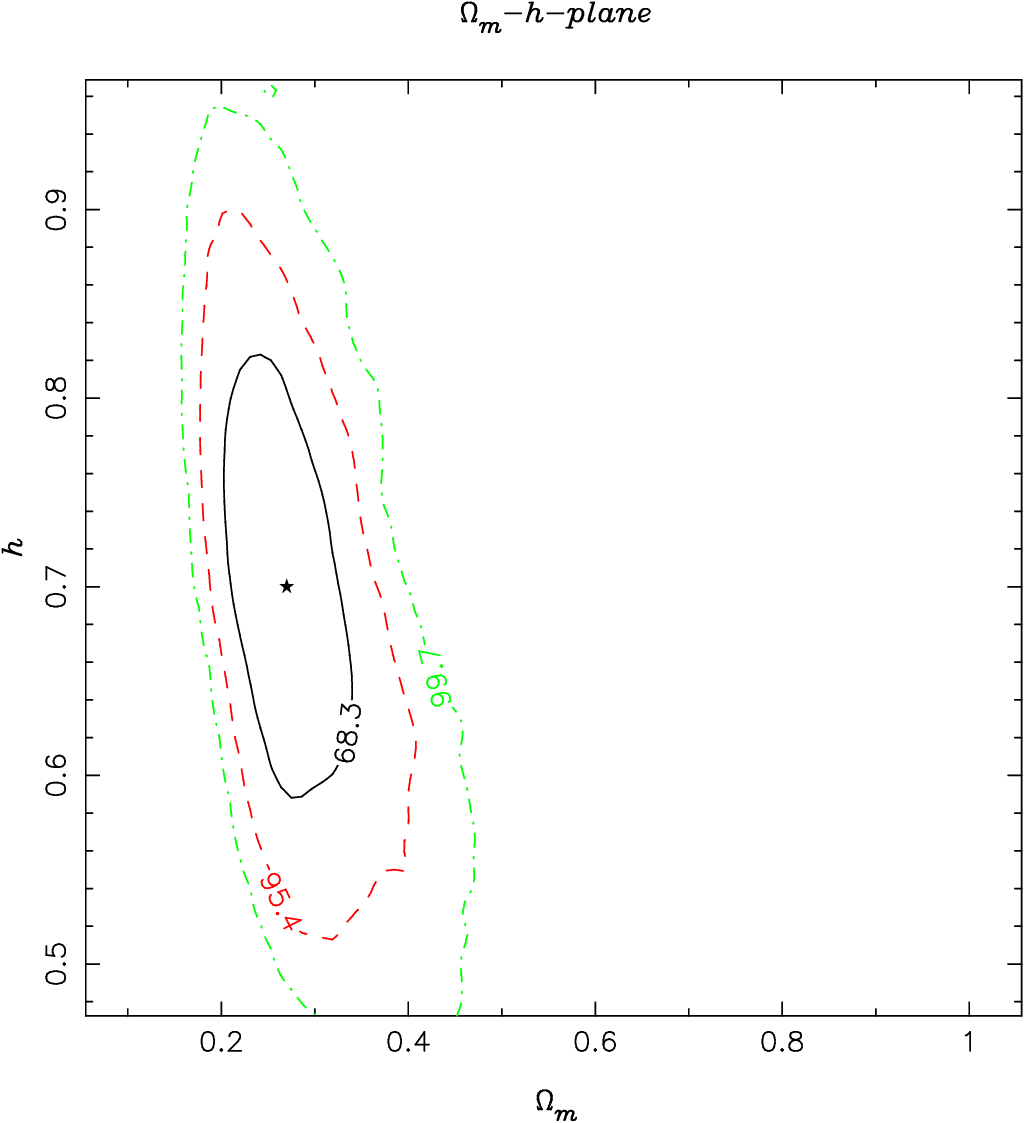}}
\subfigure{\includegraphics[angle=-90,width=0.3\textwidth]{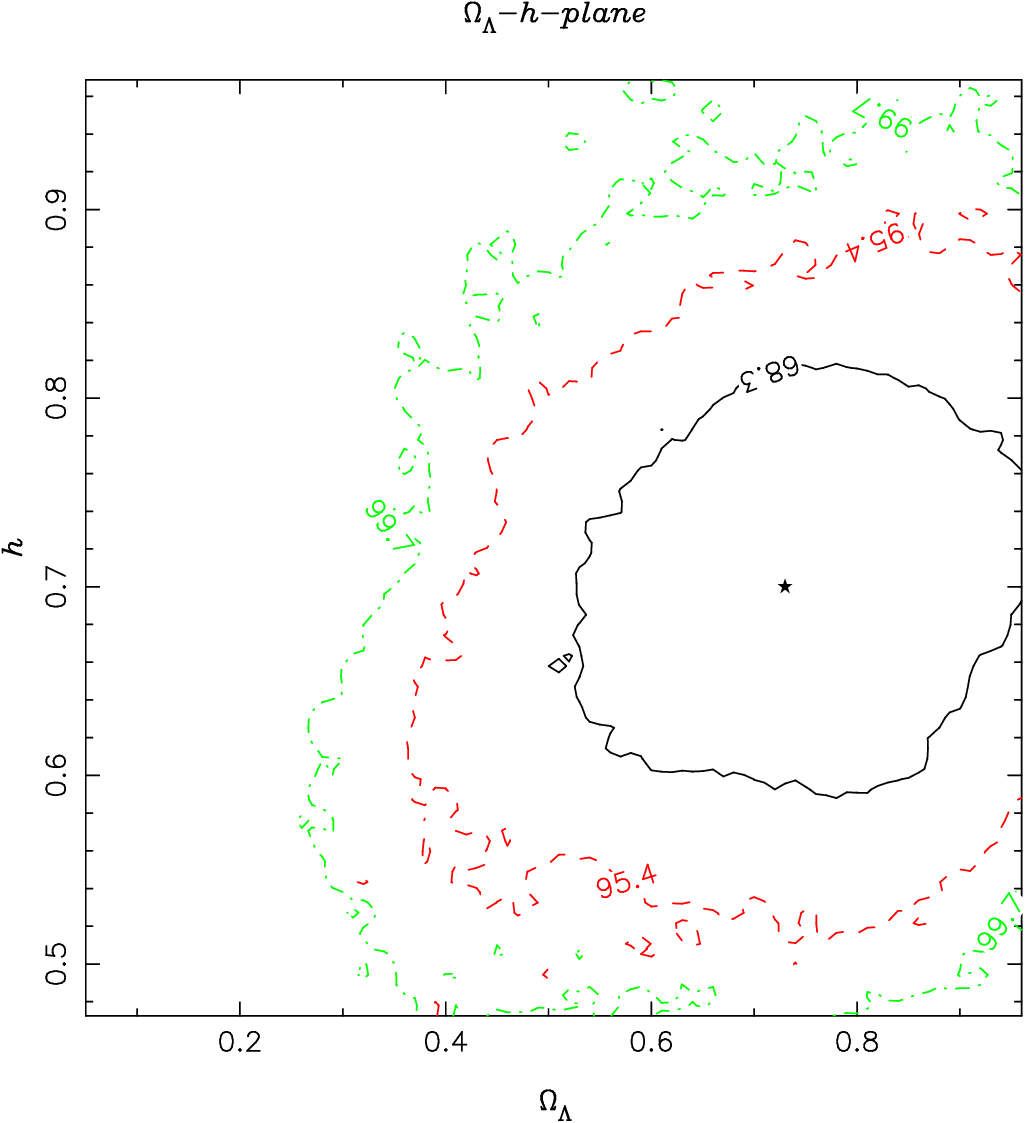}}
\subfigure{\includegraphics[angle=-90,width=0.3\textwidth]{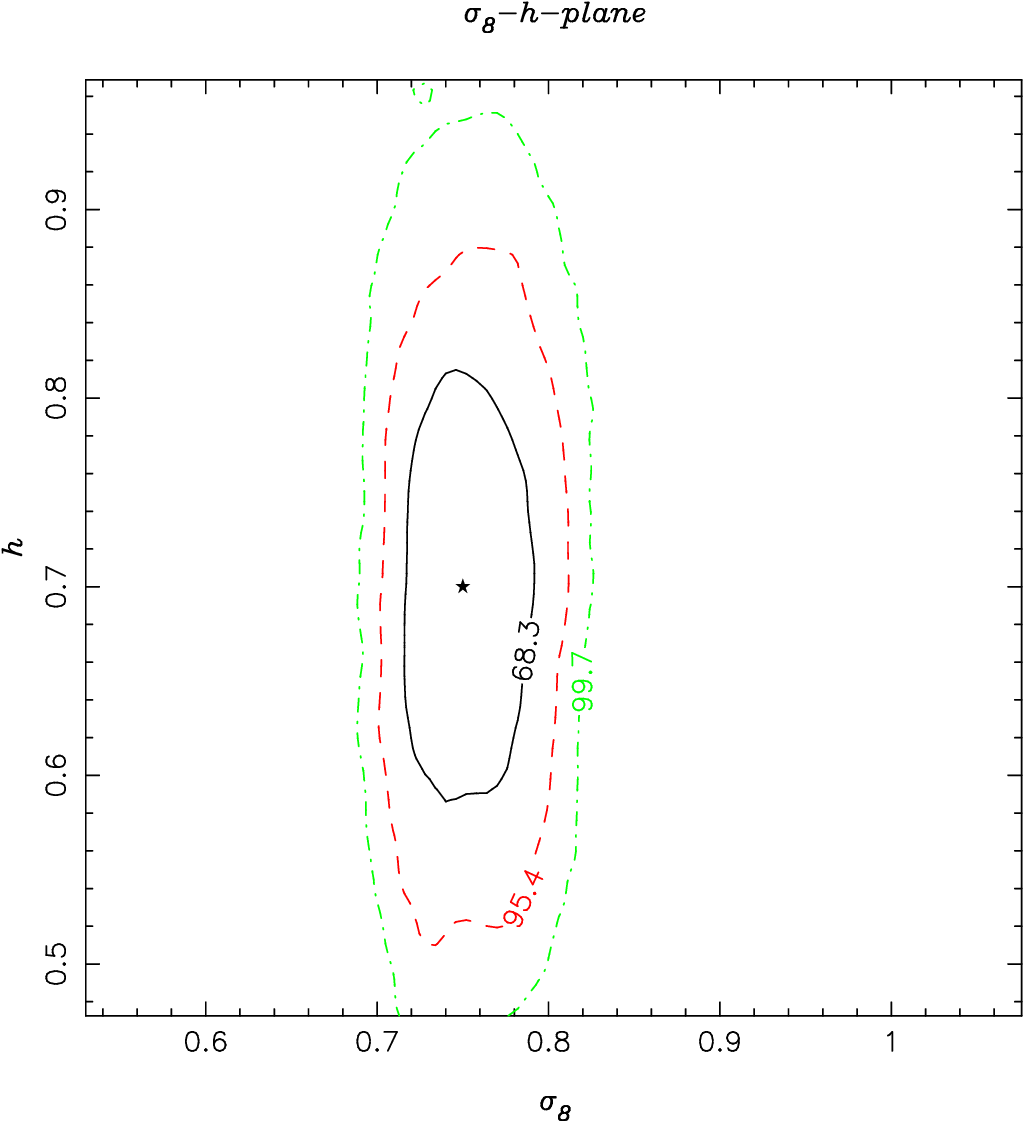}}
\caption{The same as Figure \ref{fig:2dcont} for the WMAP best
  fit fiducial cosmological parameter model ($\Omega_m=0.27$,
  $\Omega_\Lambda=0.73$, $\sigma_8=0.75$ and $h=0.7$). \label{fig:2dcontwmap}}
\end{flushleft}
\end{figure*}

The one-dimensional constraints given in Figures
\ref{fig:onedimplanckcosconstr} and \ref{fig:onedimplanckcosconstrwmap} and in
Tables \ref{tab:planckcosconstr} and \ref{tab:planckcosconstrwmap} fail to
reveal important information hidden in parameter correlations and degeneracies.
In order to display degeneracies between parameters, the two-dimensional joint
likelihood distributions for all possible pairs of parameters are shown in
Figures \ref{fig:2dcont} and \ref{fig:2dcontwmap} for our fiducial models.

The panels of Figures \ref{fig:2dcont} and \ref{fig:2dcontwmap}
display well-known degeneracies between parameters constrained by cluster
redshift number counts. For example, the $\Omega_m
- \sigma_8$ degeneracy has been found by many other authors performing an
analysis on either simulated (see e.g. \cite{2003BattyeWeller}) or observed data (see
e.g. \cite{2003BahcallBode}). Further, the shown correlation between $\Omega_m$ and
$\Omega_\Lambda$ is expected. However, since many authors restrict their analyses to
flat models, this degeneracy has been much less studied. 
The region of acceptable values of $\Omega_m$ and
$\Omega_\Lambda$ ensures that the evolution
of the growth factor of linear perturbations is approximately comparable to the one of the fiducial
cosmologies in the redshift range of interest. Large deviations in
the evolution of linear perturbation growth affect the cluster mass function
and thus the expected redshift cluster number count and its slope above a limiting mass
significantly. Moreover, the comoving
volume element and the limiting mass - both redshift dependent - as well show for the allowed
parameter combinations (region of high confidence) over the redshift range
covered by the Planck cluster sample only moderate variations from the
respective values of the fiducial models.
 From these two previous degeneracies one can predict the
$\Omega_\Lambda - \sigma_8$ one (see right panel in the second row of Figures \ref{fig:2dcont} and \ref{fig:2dcontwmap}). The Hubble parameter $h$ is degenerate to
$\Omega_m$ and $\Omega_\Lambda$. Though, for the found cluster selection it
shows little degeneracy with $\sigma_8$.

\begin{table}
\begin{center}
\begin{tabular}{||c|c||} \hline
Parameter & Median ($1\sigma$ constraint)\\
\hline
&\\
$\Omega_m$ & $0.269^{+0.047}_{-0.041}$\\
&\\
\hline
&\\
$\sigma_8$ & $0.752^{+0.026}_{-0.023}$\\
&\\
\hline
&\\
$\Omega_\Lambda$ & $0.736^{+0.132}_{-0.150}$\\
&\\
\hline
&\\
$h$ & $0.7^{+0.08}_{-0.08}$ (prior)\\
&\\
\hline
\end{tabular}
\caption{The same as Table \ref{tab:planckcosconstr}. This time giving the
  median and $1\sigma$ constraints on
  the cosmological parameters for an underlying cosmological model with WMAP
  best-fit parameters. \label{tab:planckcosconstrwmap}}
\end{center}
\end{table}

Note that evidence of the multi-dimensionality of the degeneracies becomes
noticeable by comparison of Figure \ref{fig:2dOmsig8concordwmap} with the left panel in the second row
($\Omega_m$-$\sigma_8$-plane) of Figure \ref{fig:2dcont} and Figure \ref{fig:2dcontwmap} respectively. The tight
prior on $h$ carves out regions around the parameters of each fiducial model to which
the constraints are confined. These regions in
the two-dimensional parameter space are localised within the two-dimensional
constraints plotted in Figure \ref{fig:2dOmsig8concordwmap}. Shifting the mean of the Gaussian prior to
a lower value of $h$ shifts the region of high confidence further to the right
on the $\Omega_m$-axis and marginally downwards on the $\sigma_8$-axis. The
opposite happens for an increase of the mean value of the Gaussian Hubble
parameter prior (while keeping the fiducial parameters unaltered). This illustrates the strong degeneracy between the matter
density $\Omega_m$ and the Hubble parameter $h$.
Therefore, constraining $h$ leads to a strong enhancement of the constraint on
the matter density and slightly improves the constraint on $\sigma_8$.

\section{Discussion and conclusions}
\label{sec:plmfmfmcmcconcl}

The sky simulation(s) and the modelling of the observing process of the Planck
Surveyor satellite presented in this work are of high realism and are based on
recent observational
constraints and predictions obtained from numerical simulations. 
Nevertheless, as discussed previously, some uncertainties concerning the component
modelling remain. These might even have more than a marginal influence on the
results. For example, the modelling of the number counts of IR/SM point
sources and their spatial correlation to galaxy clusters is speculative
since available observational data are sparse. Apart from a few small patch
observations undertaken by SCUBA and MAMBO (see e.g. \cite{2004Greveetal}),
there is little known about the point source population at submillimetre and millimetre
wavelengths, resulting in high sample variance and apparently hardly any insight into correlations.
Moreover, some channels of the Planck HFI (100 GHz, 150 GHz and 353 GHz) to which IR/SM
sources contribute are highly valuable for SZ cluster detections. Therefore, a
higher level of point source contamination at these frequencies and/or them
being (strongly) spatially correlated with clusters could affect Planck cluster
number counts. Furthermore, cluster internal physical
processes, such as AGN or
SN cluster gas heating, may contribute to the cluster SZ signal in addition to
gravitational processes. So far, the mechanisms of these processes occuring at
late cluster evolution stages are not well understood. However, the Planck
cluster sample should provide an extensive basis for studying such cluster physics.

To the simulated data we have applied a cluster extraction algorithm. 
 The method is a multi-frequency matched filtering
technique. It is based on a variational cluster template whose parameters are
discretely varied. For this algorithm we optimised the parameter
discretisation with respect to algorithm performance and computing cost. 
 Contrary to past analyses which have often restricted the template to be rigid
(e.g. Gaussian beam shaped under the assumption that sources are unresolved), the priors on
the template parameters have been chosen in such a way that they yield an optimisation of the
cluster detection efficiency for the expected quality of the Planck data and
cluster physical sizes. The recovered cluster catalogue is then constructed from
candidates whose detection significance exceeds $5\sigma$.
The built cluster catalogue has been found to be suitable for
  cosmological considerations under the condition that the survey selection
  for the data and the adopted algorithm is
  well understood.
\footnote{This assumption has been made
  throughout this work.} For a suitable parameterised selection function,
  a `self-calibration' is as well feasable due to the large sample size. The
  extracted catalogue consist of clusters at moderate and intermediate
  redshifts with cluster masses generally of $M_{\rm cl} \gtrsim 5\times
  10^{14} h^{-1}M_{\odot}$. 
 The contamination of the catalogue
  is found to be fairly low. 
 In general, an expected low sample
  contamination is a prerequisite in order to be able to use reliably the
  statistical power of such a large catalogue in addition to a comprehensive
  sample completeness. 

Furthermore, comparing this work with the work presented in
GKH05, it is found that the cluster selection is in
good agreement with the one found in our previous paper (even
though cluster extraction methods differ).  In
GKH05, due to the strict matching acceptance region
of $\sim 2$ arcmins, a certain
number of clusters ($\lesssim 10$ per cent above a flux limit of $2\times 10^{-3}$
arcmin$^{2}$) are not matched up correctly. These missed matches reduce the
sample completeness estimate and increase on the other hand the contamination
estimate by approximately the same percentage. However,
our previous work aimed to give conservative estimates and reliable limits
which should be definitely achievable by the Planck cluster
sample. On the contrary, overly
large matching acceptance regions 
lead to over-estimates of the completeness
and purity above survey flux detection thresholds. 
This is especially the case for low flux limits and high cluster surface
densities, for which overly large acceptance regions increase highly 
 the possibility of finding a match just by chance.

The precision of the photometric
 cluster parameters of the Planck samples recovered by the applied algorithms
 is (on the basis of our simulations) expected to be rather poor. This is
 similar to what we found in GKH05 using a different
 cluster recovery pipeline. Here, the (relative) dispersion ($\sigma(\log Y)$)
 of the reconstructed cluster fluxes around their real values is estimated to
 be approximately 15 percent for the whole sample on average for the recovery
 by the MFMF method.
 Though, it is found that the photometric accuracy 
 improves with
 increasing sample flux threshold. Nevertheless, we have not made use of
 recovered photometric cluster properties (namely the recovered cluster
 fluxes) in our parameter analysis. 
The found large dispersion 
 reduces the 
 usefulness of the recovered cluster fluxes 
 for
 survey `self-calibration' and cosmological parameter constraints via
 accurate theoretical mass function predictions, a parameterised mass observable relation and
 selection function.
Likewise, the low photometric quality of the Planck cluster sample affects its cluster physical
interpretation 
 in the exact same manner. 
Therefore, we only touched briefly aspects of late cluster physics in our
discussion above. 
 Only global
 trends, such as the overall (average) normalisation of the $M-Y$ scaling relation, may
 be grasped 
 by the Planck sample (see GKH05). Constraints on the scaling relation normalisation, however, are expected to be degenerate with
 cosmological parameters, such as $\sigma_8$.
One way to improve the understanding of cluster physics is to 
follow up the sample clusters with observations in the optical and X-ray wavebands. 
 As discussed in section \ref{sec:expeccosconstrplanck}, this is a rather cumbersome undertaking.
It further has to be pointed out that a template choice differing from the actual
 universal one biases the photometric parameter estimates (on average) in addition to
the large dispersion. 
 However, at the time Planck completes its data collection, several `small scale' SZ experiments will
have finished their scientific programmes and obtained results which will give
insights into cluster physical aspects, such as cluster profiles, the normalisation of the $M-Y$
relation and its evolution and intrinsic scatter. Therefore, our assumption of available prior
information is realistic.
These experiments can also be used to follow up the Planck sample in the
microwave band enhancing resolution and cluster flux estimations.

Besides optimising cluster extraction and forecasting 
 cluster selection of two
powerful algorithms applied to simulated Planck data, we focus on the
cosmological prospects which can be 
 accomplished by the Planck cluster sample. 
This is the first time that based on a realistic selection function derived
from astrophysical observation simulations and an implemented data analysis pipeline the cosmological
use of the future Planck cluster survey is evaluated.
In our MCMC analysis to constrain cosmological parameters, we assume a priori
knowledge about the mass observable relation and the cluster selection. As
 pointed out in the previous paragraph, limits on the $M-Y$ relation
 normalisation will be
 obtainable before long by
up-comming SZ cluster survey instruments and follow-up of their observations. 
 Insights into the cluster selection function can be achieved by
mock observations as presented in this work. However, neglecting or misestimating
the magnitude of contaminants leads inevitably to a bias in the
expected selection and therefore in the best-fit
cosmological parameters obtained from the sample. 
 Note that even in the case of a so-called `self-calibration'
analysis some advanced fixings of the parameterised (functional) shape of the selection
have to be made. For example, a
very common assumption is that the 
 scatter about the mean limiting
mass or value of the observable respectively
is 
 of Gaussian nature.\footnote{This assumes
that the intrinsic scatter in the mass-observable relation and the (in
quadrature) additive (extra)
scatter in the reconstructed cluster fluxes caused by contamination are Gaussian.} 
 Apart from mock simulations as performed in this work to estimate the selection
function, cross-checks of simulations with data and direct contaminant
extraction from data are essential to investigate cluster selection. 
 Reliably pinning down the selection function of the sample 
will be an iterative process in which results from mock simulations 
will have to be adjusted to observations in order to make them converge.
A number of methods are available to seperate spatially and spectrally diverse
components and thus help with estimating the flux selection of clusters. 
 Powerful algorithms for disentangling components are, for example, Independent Component
Analysis (ICA) and Maximum Entropy methods (MEM). They may also be applied in
order to 
 reduce contamination. However, it
is as well possible to estimate the confusion of the cluster fluxes 
 by the MFMF cluster extraction algorithm
itself by allowing the level of contamination to vary and including it in
the parameter optimisation. 

Before summarising the results on constraints, some further challenges
the parameter estimation will have to face are outlined. 
Firstly, uncertainties in cluster physics as well as in cosmological parameters and
foregrounds (e.g. IR/SM point sources) suggest
that the cluster number contained in the future Planck sample may be uncertain up to
a factor of 2-3. The number of recovered clusters evidently controls the statistical ability of
the sample to tighten confidence limits on the best-fit parameters.
Hence, in section \ref{sec:expeccosconstrplanck} the impact of different the
Universe possibly underlying
cosmologies 
 on the ability to derive tight cosmological parameter constraints from the
 respective Planck cluster sample has been
 investigated.
The two different fiducial cosmologies utilised are 
 the concordance $\Lambda$CDM and the best-fit $\Lambda$CDM WMAP
model. 
They differ mainly in their value of $\sigma_8$, a parameter which is still
fairly little constrained (by today's standards).
Unsurprisingly, the sample with the fewer cluster members (lower value of $\sigma_8$) places the weaker
constraints on the parameters (about a factor of 1.8 for the $\sigma_8$
constraint; see Tables \ref{tab:planckcosconstr} and \ref{tab:planckcosconstrwmap} and Figures \ref{fig:onedimplanckcosconstr} and \ref{fig:onedimplanckcosconstrwmap}).
Another concern which is not linked (directly) to the recovery process is the
level of accuracy in the theoretical prediction of the cluster mass
function. Comparison of large scale numerical (N-body) simulations and halo
finding codes estimate the current theoretical uncertainty to be at a level of
approximately 10 percent. 

Disregarding the theoretical mass function uncertainty, our forecasts suggest that the Planck cluster sample will be
able to put tight constraints on cosmological parameters additional to the
ones derived from the primordial CMB power spectrum recovered from Planck
data. Despite the sample's rather `low' expected mean redshift, due to the full
sky coverage of the survey, cosmological parameter constraints of
similar quality as gained from current primordial CMB analyses are realisable.  The Planck cluster survey will especially have the
capability to tighten constraints on $\sigma_8$.
The matter variance on scales of 8 $h^{-1}$Mpc is a parameter that can be only
`weakly' (in comparison to other parameters) 
 constrained by primordial CMB measurements.
All over, current primordial CMB observations on their own are not
best suited to place constraints on the shape and normalisation of the
matter power spectrum.
The large difference of the best-fit $\sigma_8$ value of the first and three
year WMAP data (which show an approximately $2\sigma$ discrepancy) indicates
that the current primordial CMB parameter constraint should possibly not be taken
too literally. This notion is strengthened by the fact that several other
experiments obtained values discrepant to the WMAP constraint.  Large scale structure observations of galaxies and clusters are
more suited to place tight constraints on the shape of the matter power
spectrum 
 and $\sigma_8$. However, most of the large scale structure
surveys carried out up to the present day are of such small scale that they
are heavily affected by sample variance and ignored systematics may play a
role as well. The Planck cluster sample can overcome these problems.
Furthermore, degeneracies between parameters, such as $\Omega_m$ and
$\Omega_\Lambda$, obtained from primordial CMB measurements and cluster number counts
are different. Therefore performing a combined data analysis helps to narrow down regions of high likelihood in
parameter space and to break the parameter degeneracies. 
 Further, 
 we find that while the constraint on $\sigma_8$ is only
weakly dependent 
 on a prior on the Hubble
  parameter, constraints on the matter and dark energy density, $\Omega_m$ and $\Omega_\Lambda$,
  depend strongly on it. By including a reasonable prior on $h$ in the
  analysis as given, for example, by the HST Key Project, the
  confidence intervals of these parameters
 shrink by up to an order of magnitude. Nevertheless, even
  without such $h$ prior, the bare existence of a cosmological
  constant 
 can be confirmed well above the 3$\sigma$ confidence level.

In conclusion, the gain of the Planck cluster catalogue will be twofold. Firstly, it will
be a fruitful sample to serve as a base for 
studying cluster physics (the normalisation, evolution and intrinsic scatter
of cluster properties and their scaling relations) in combination with large scale surveys at other
wavelengths (RASS, SDSS etc.) and/or follow-up in the microwave band. 
 Secondly, our investigations suggest that the Planck cluster sample
 (recoverable from future Planck data by algorithms like the one described above) can
 live up to the high expectations predicted from pure
theoretical and analytical estimations by placing meaningful constraints on
cosmological parameters. 
Overall, an all-sky sample of massive clusters with a well understood
selection function as achievable by the Planck mission will be of great value
for cluster research and cosmology. In a forthcoming paper we will investigate
more sophisticated methods than MFMF to examine how their performances improve
cluster extraction and constraints on cosmological parameters. 

\section{Acknowledgements}

JG acknowledges an Isaac Newton scholarship from the Cambridge
University Isaac Newton Trust and funding from the Cambridge Philosophical Society.
JG is grateful to Jim Bartlett and the French Research Ministry for a
postdoctoral fellowship. 
 JG
thanks the Group PCC/APC formerly at College de France and the head of the
Group, Yannick Giraud-Heraud, and the Radio Astronomy Group at the Argelander
Institute for Astronomy (University of Bonn) and its head, Frank Bertoldi, for their hospitality.
Furthermore, JG acknowledges helpful conversations with Vlad Stolyarov and Mark Ashdown
about instrumental properties of the Planck Surveyor satellite. 

\bibliographystyle{mn2e}
\bibliography{paper}

\label{lastpage}

\end{document}